\documentclass[twocolumn]{IEEEtran}
\usepackage{amssymb}
\usepackage{amsmath}
\usepackage[version=4]{mhchem}
\usepackage{graphicx}
\usepackage{multirow}
\usepackage{wrapfig}
\usepackage{color}
\usepackage{relsize}
\usepackage{colortbl}
\usepackage{kbordermatrix}
\usepackage{bm}
\usepackage{soul}
\usepackage{subfigure}
\usepackage{enumerate}
\usepackage{paralist}
\usepackage{dsfont}
\usepackage[normalem]{ulem}
\usepackage{mathtools}
\usepackage{tabularx,booktabs,caption}
\usepackage[flushleft]{threeparttable}\usepackage{tabularx,booktabs,caption}
\usepackage[flushleft]{threeparttable}

\newcommand{\p}{\mathrm{p}}
\newcommand{\B}{\mathrm{B}}
\newcommand{\U}{\mathrm{U}}

\newcommand{\E}{\mathrm{E}}
\newcommand{\e}{\mathrm{e}}
\newcommand{\Var}{\mathrm{Var}}
\newcommand{\Std}{\mathrm{Std}}

\newcommand{\Cov}{\mathrm{Cov}}

\def\mathclap#1{\text{\hbox to 0pt{\hss$\mathsurround=0pt#1$\hss}}}
\makeatletter
\newcommand{\vastt}{\bBigg@{3}}
\newcommand{\vast}{\bBigg@{4}}
\newcommand{\Vast}{\bBigg@{5}}
\newcounter{mytempeqncnt}
\makeatother
\DeclareMathOperator*{\argmax}{arg\,max}

\DeclareMathOperator\erfc{erfc}

\DeclarePairedDelimiter\floor{\lfloor}{\rfloor}

\begin{document}
\title{Detection in Molecular Communications with Ligand Receptors under Molecular Interference}
\author{Murat Kuscu,~\IEEEmembership{Member,~IEEE}
        and Ozgur B. Akan,~\IEEEmembership{Fellow,~IEEE}
       \thanks{An earlier version of this work was presented at IEEE SPAWC'18, Kalamata, Greece \cite{muzio2018selective}. }
       \thanks{The authors are with the Department of Electrical and Electronics Engineering, Koc University, Istanbul, 34450, Turkey  (email: \{mkuscu, akan\}@ku.edu.tr).}
       \thanks{Ozgur B. Akan is also with the Internet of Everything (IoE) Group, Electrical Engineering Division, Department of Engineering, University of Cambridge, Cambridge, CB3 0FA, UK (email: oba21@cam.ac.uk).}
\thanks{This work was supported in part by the ERC (Project MINERVA, ERC-2013-CoG \#616922) and by the AXA Research Fund (AXA Chair for Internet of Everything at Koc University).}}


\maketitle

\begin{abstract}
	Molecular Communications (MC) is a bio-inspired communication technique that uses molecules to transfer information among bio-nano devices. In this paper, we focus on the detection problem for biological MC receivers employing ligand receptors to infer the transmitted messages encoded into the concentration of molecules, i.e., ligands. In practice, receptors are not ideally selective against target ligands, and in physiological environments, they can interact with multiple types of ligands at different reaction rates depending on their binding affinity. This molecular cross-talk can cause a substantial interference on MC. Here we consider a particular scenario, where there is non-negligible concentration of interferer molecules in the channel, which have similar receptor-binding characteristics with the information molecules, and the receiver employs single type of receptors. We investigate the performance of four different detection methods, which make use of different statistics of the ligand-receptor binding reactions: instantaneous number of bound receptors, unbound time durations of receptors, bound time durations of receptors, and combination of unbound and bound time durations of receptors within a sampling time interval. The performance of the introduced detection methods are evaluated in terms of bit error probability for varying strength of molecular interference, similarity between information and interferer molecules, number of receptors, and received concentration difference between bit-0 and bit-1 transmissions. We propose synthetic receptor designs that can convert the required receptor statistics to the concentration of intracellular molecules, and chemical reaction networks that can chemically perform the computations required for detection. 
	
\end{abstract}


\IEEEpeerreviewmaketitle

\section{Introduction}
\label{sec:Introduction}
Internet of Bio-Nano Things (IoBNT) is an emerging technology built upon the artificial heterogeneous communication networks of nanomachines and biological entities, promising for novel applications such as smart drug delivery with single-molecule precision and continuous health monitoring \cite{akyildiz2015internet, kuscu2016internet, dinc2019internet, akyildiz2019microbiome}. Bio-inspired Molecular Communication (MC) has emerged as the most promising communication technique to enable IoBNT applications. MC uses molecules, instead of electromagnetic waves (EM), to transfer information, which can be encoded into the concentration or type of molecules \cite{atakan2016molecular, akan2017fundamentals, akyildiz2019moving}. Being fundamentally different from conventional EM communication techniques, MC has brought about new interdisciplinary challenges in developing communication techniques and transceiver architectures.




Many efforts in MC research have been devoted to developing channel models and low-complexity communication techniques \cite{akyildiz2019moving, jamali2019channel, kuscu2011physical, kuscu2019transmitter, farsad2016comprehensive}. Of particular interest has been the detection problem. Several detection methods of varying complexity have been developed for different device architectures \cite{kuscu2019transmitter, kilinc2013receiver, li2019csi, li2015low}. Most studies focusing on MC detection, however, consider a particular receiver architecture that is capable of counting every single molecule inside its virtual reception space \cite{pierobon2011diffusion, kuscu2019transmitter}. On the other hand, an increasing research interest is being directed towards MC receivers with ligand receptors, which chemically interact with information molecules through ligand-receptor binding reaction \cite{pierobon2011noise, chou2015maximum, kuscu2018modeling, kuscu2016physical, kuscu2016modeling}. This receiver design is the most physically relevant, as the ligand-receptor interactions are prevalent in biological systems, and thus suitable for synthetic biology-enabled MC device and system architectures \cite{bialek2012biophysics, soldner2019survey, unluturk2015genetically}. This additional layer of biological interaction, while adding to the complexity of the overall MC system, yields interesting statistics that can be exploited in order to develop reliable detection methods.


MC detection with ligand receptors has been widely studied; however, in existing studies, receptors are assumed to be ideally selective against the information molecules \cite{kuscu2019transmitter}. On the other hand, in practice, the selectivity of biological ligand receptors is not ideal, and receptors can bind other types of molecules that have a nonzero affinity with the receptors.
This molecular interference, also called cross-talk, is widely observed in various biological systems due to the prevalence of promiscuous ligand-receptor interactions \cite{lalanne2015chemodetection, nobeli2009protein}. A paradigmatic example is the immune recognition where T cells express promiscuous T cell receptors (TCRs) that bind both self-ligands and a large spectrum of foreign ligands \cite{sewell2012must, baker2012structural}. The detection of foreign ligands via TCR signaling evokes the immune response. Other examples include the transcriptional cross-talk due to nonspecific binding of genes and regulators in gene regulation \cite{oeckinghaus2011crosstalk}, quorum sensing (QS) where QS receptors are promiscuously activated by multiple types of ligands \cite{wellington2019quorum}, and most of the cellular communication systems, such as bone morphogenetic protein (BMP), Wnt, Notch, and fibroblast growth factor (FGF) signaling pathways \cite{su2020ligand}.


At the heart of the widespread cross-talk in biological systems lies the promiscuous proteins including cellular receptors, enzymes and antibodies, that can interact with diverse ligand structures, including small molecules, macromolecules, and ions \cite{nobeli2009protein, tawfik2010enzyme, jain2019antibody}. The prevalence of promiscuous protein interactions is mainly due to the degeneracy of the protein interface structures \cite{gao2010structural}. This can be further attributed to the structural flexibility of proteins and interacting ligands, partial recognition due to the tolerance in shape and chemical complementarity in binding, and the presence of multiple interaction sites in proteins \cite{nobeli2009protein}.


%

Many biological processes, such as T cell antigen recognition and transcription/translation, are adapted to preserve the specificity in the presence of cross-talk through complex out-of-equilibrium biophysical mechanisms, such as kinetic proofreading (KPR) and adaptive sorting \cite{yousefi2019optogenetic, lalanne2013principles}. Some sensory systems, such as odor recognition system, even seem to exploit the cross-talk as an opportunity to expand the ligand search space with a limited number of receptors \cite{hallem2006coding, carballo2019receptor}. In bacteria QS, the cross-talk is considered as a mechanism that brings evolutionary advantage by facilitating the interspecies interactions in complex microbial communities \cite{wellington2019quorum}. During cell-cell communication in multicellular organisms, promiscuity of ligand-receptor interactions is shown to enable individual cells to address a larger number of cell types through combinatorial addressing \cite{su2020ligand}.   


Considering that the synthetic biological MC systems will utilize receptors and ligands derived from biological systems, the promiscuity of the natural ligand-receptor interactions is likely to be preserved in MC devices. This could result in significant level of molecular interference in biologically relevant environments, e.g., inside human body, which are typically crowded with diverse types of proteins (e.g., transcription factors, enzymes, hormones), other macromolecules (e.g., nucleic acids), and small molecules, that can promiscuously bind the receptors synthesized on the MC receiver surface \cite{uhlen2015tissue, thul2017subcellular, luck2020reference}. Moreover, the problem can be translated into multi-user interference in crowded multi-user MC networks, where structurally similar molecules from the same family, such as isomers, are likely to be used to enable molecule-division multiple access without increasing the burden on MC transmitters and receivers \cite{dinc2017theoretical, kim2013novel}. 





In this paper, we study the effect of molecular interference due to nonspecific ligand-receptor interactions on the reliability of synthetic MC systems. We consider an MC system encoding binary information into the concentration of molecules, i.e., utilizing binary concentration shift keying (binary CSK). The interference is resulting from a second type of molecule existing in the MC channel, whose number in the receiver's vicinity at the time of sampling follows Poisson distribution. Under these conditions, we investigate the performance of four different detection methods, which exploit different statistics of the ligand-receptor binding reaction. 

The first detection method relies on the number of bound receptors at the sampling time, which gives a measure of the total molecular concentration around the receiver. This method is the most widely studied one in MC literature concerning ligand receptors \cite{einolghozati2011capacity}.  The second method uses the maximum likelihood (ML) estimate of the total ligand concentration based on the receptors' unbound time intervals. This method has been previously introduced to overcome the saturation problem of receptors exposed to a high concentration of molecules \cite{kuscu2018maximum}. The third method relies on the estimation of the concentration ratio of information molecules, i.e., the ratio of information molecule concentration to total molecular concentration in the vicinity of the receiver, based on receptors' bound time intervals. This technique exploits the difference in the receptor binding affinities of information and interferer molecules reflected to the expected sojourn time of receptors in the bound state. The last method combines the estimates of the total ligand concentration and the concentration ratio of information molecules to obtain an estimate of the individual concentration of information molecules, which is then used for detection of molecular messages. This technique utilizes both the unbound and bound time intervals of the receptors. 

We derive the bit error probability (BEP) for each detection method, which is then used for comparing the performances of the introduced detection methods for varying strength of molecular interference, similarity between interferer and information molecules in terms of their affinity against receptors, number of receptors, and difference in received concentrations of information molecules for bit-0 and bit-1 transmissions. 
We also provide a comprehensive discussion on the practical implementation of these detection methods by biosynthetic devices. In particular, we propose synthetic receptor designs for the transduction of required receptor statistics, i.e., number of bound receptors, receptor bound and unbound time intervals,  into the concentration of intracellular molecules. We also propose a chemical reaction network (CRN) for each method that can perform the analog and digital computations required for detection.

The remainder of the paper is organized as follows. In Section \ref{sec:related}, we provide a brief overview of the related work. In Section \ref{sec:statistics}, we review the fundamentals of ligand-receptor binding reactions. The considered MC system is described in Section \ref{sec:system} along with accompanying assumptions.  We introduce the detection methods in Section \ref{sec:detection}, where we also derive the BEP for each detection method. The results of the performance evaluation are discussed in Section \ref{sec:performance}. In Section \ref{sec:implementation}, we provide a discussion on the implementation, and propose synthetic receptor designs for signal transduction and CRNs for intracellular calculations. Finally, we conclude the paper in Section \ref{sec:conclusion}.





\section{Related Work}
\label{sec:related}
MC detection has been extensively studied for various modulation schemes, channel types, and receiver architectures. In developing received signal models for detection, simplifying assumptions are often utilized to address the intricate physical interaction between the channel and the receiver. Most of the existing MC detection studies assume passive and transparent receivers, and ignore the complex ligand-receptor interactions. Many other studies consider an absorbing architecture that actually corresponds to a receiver with infinite number of receptors, which can irreversibly react with information molecules and absorb them, i.e., remove them from the channel, at infinitely high reaction rates. Although passive and absorbing receiver architectures have little correspondence with reality, these studies provide useful theoretical performance bounds for MC detection. A recent comprehensive review of the existing MC detection methods can be found in \cite{kuscu2019transmitter}. 

The interest in MC detection with ligand receptors, on the other hand, has only recently gained momentum. In \cite{chou2013extended, chou2015markovian}, a modeling framework based on CMTPs has been introduced. Using this framework, Maximum A Posteriori (MAP) decoders have been developed based on sampling the continuous history of the receptor states, including the exact time instances of the binding events \cite{chou2015maximum, chou2019designing}. In \cite{kuscu2018maximum}, we proposed an ML detection method based on receptor unbound times to overcome the saturation problem of reactive receivers with finite number of ligand receptors. However, none of these studies consider the existence of similar types of molecules interfering with the ligand-receptor binding reaction. 

There is also a substantial body of work in biophysics literature concerning the theoretical bounds of molecular sensing with ligand receptors. Regarding the interference on the ligand-receptor reactions, in \cite{mora2015physical, singh2017simple, lalanne2015chemodetection, siggia2013decisions}, authors investigate the ML estimation of the concentrations of two different ligand types based on receptor bound times. These studies also suggest that certain types of living cells, e.g., T cells in the immune system, might be implementing similar estimation methods in discriminating against the foreign agents through a KPR mechanism, in which receptors sequentially visit a number of internal states to sample the duration of binding events. Following a similar approach with these studies, in \cite{kuscu2019channel}, we introduced a novel channel sensing method that can concurrently estimate the concentration of several different types of ligands using the receptor unbound and bound times. Lastly, in an earlier version of this study \cite{muzio2018selective}, we investigated the theoretical performance bounds of ML detection based on receptor bound times, instantaneous number of bound receptors, and receptor unbound times. 

Different from the conference version \cite{muzio2018selective}, this study investigates four practical detection methods that can be implemented by biological MC receivers with the use of synthetic receptors and CRNs. We derive analytical expressions for bit error probability, and propose synthetic receptor designs and CRNs for sampling the receptor states and performing the detection by biochemical means.  

\section{Statistics of Ligand-Receptor Binding Reactions}
\label{sec:statistics}

In ligand-receptor binding reaction, receptors randomly bind external molecules, i.e., ligands, in their vicinity. Following the canonical Berg-Purcell scheme, the stochastic ligand-receptor binding process can be abstracted as a continuous-time Markov process (CTMP) with two states; corresponding to the bound (B) and unbound (U) states of the receptors \cite{berg1977physics, ten2016fundamental}. Due to the memoryless property of the Markov processes, the dwell time at each receptor state follows exponential distribution \cite{liggett2010continuous}, with a rate parameter depending on the kinetic rate constants of the ligand-receptor binding reaction given as    
\begin{equation}
\ce{U  <=>[{c_L(t) k^+}][{k^-}] B},
\label{eq:bindingreaction}
\end{equation}
where $c_L(t)$ is the time-varying ligand concentration, $k^+$ and $k^-$ are the binding and unbinding rates of the ligand-receptor pair, respectively \cite{berezhkovskii2013effect}. The correlation time of this Markov process, which can be regarded as the relaxation time of the ligand-receptor binding reaction to equilibrium, is given by $\tau_B = 1/\left(c_L(t) k^+ + k^-\right)$ \cite{berezhkovskii2013effect, ten2016fundamental}. In diffusion-based MC, due to the low-pass characteristics of the diffusion channel, the bandwidth of the $c_L(t)$ is typically significantly lower than the characteristic frequency of the binding reaction \cite{pierobon2011noise}, which is given by the reciprocal of the receptor correlation time i.e., $f_\text{B} = 1/\tau_B = c_L(t) k^+ + k^-$. Therefore, the receptors are often assumed to be at equilibrium with a stationary ligand concentration, which is hereafter simply denoted by $c_L$. At equilibrium, the ligand-receptor binding reaction obeys the detailed balance, such that the rate of unbinding transitions must be equal to the rate of binding transitions, i.e., $\p_\B k^- = (1-\p_\B) c_L k^+$ \cite{endres2009maximum}. Here, $\p_\B$ is the probability of finding a receptor at the bound state at equilibrium, which can be obtained from the detailed balance condition as 
\begin{equation}
	\p_\B  = \frac{c_L k^+}{c_L k^+ + k^-} = \frac{c_L}{c_L+K_D},
	\end{equation}
where $K_D = k^-/k^+$  is the dissociation constant, which gives a measure of the \textit{affinity} between a ligand and a receptor. 
\begin{figure}[!t]
	\centering
	\includegraphics[width=9cm]{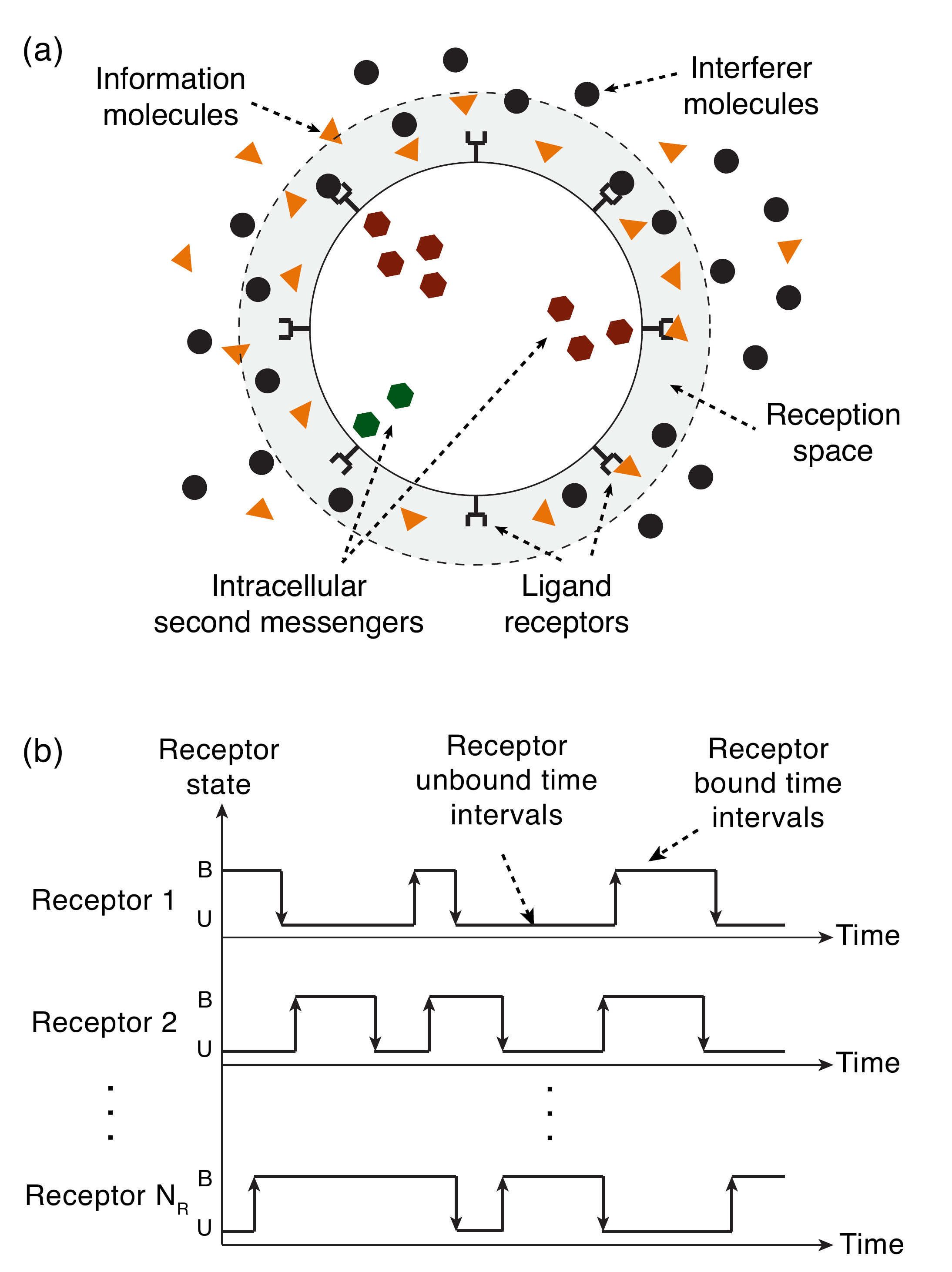}
	\caption{(a) A biological MC receiver with ligand receptors which can bind both information and interferer molecules. Receptors encode one or more statistics of the binding events into the concentration of intracellular molecules. (b) An example time trajectory of receptors fluctuating between the bound and unbound states. }
	\label{fig:interference}
\end{figure}


In the presence of two different types of molecules, e.g., information and interferer molecules, in the channel, as shown in Fig. \ref{fig:interference}(a), which can bind the same receptors with different affinities, i.e., with different dissociation constants, the bound state probability of a receptor at equilibrium becomes 
\begin{equation} \label{probBinding2}
\p_\B = \frac{c_s/K_D^s + c_{in}/K_D^{in}  }{1 + c_s/K_D^s + c_{in}/K_D^{in}},
\end{equation}
where $c_s$ and $c_{in}$ are the concentrations of information and interferer molecules, whose dissociation constants are denoted by $K_D^s$ and $K_D^{in}$, respectively (please refer to Appendix \ref{AppendixA} for the derivation). If the receiver has $N_R > 1$ independent receptors, the number of bound ones at equilibrium follows binomial distribution with the number of trials $N_R$ and the success probability $\p_\B$. 


On the other hand, the duration for which the receptors stay bound or unbound can reveal more information about the concentration and type of the molecules co-existing in the channel \cite{kuscu2019channel}. The likelihood of observing a set of $N$ independent binding and unbinding time intervals over any set of receptors at equilibrium can be written as
\begin{align}\label{eq:likelihood1}
\p\left(\{\tau_b,\tau_u \}_{N}\right) &= \frac{1}{Z} \e^{-\mathlarger{\sum_{i=1}^N} \tau_{u,i} \left( k_s^+ c_s + k_{in}^+ c_{in} \right)}  \\ \nonumber
& \times \mathlarger{\prod_{i=1}^{N}}  \left(\sum_{j=\{s, in\}} k_j^+ c_j k_j^- \e^{-k_j^- \tau_{b,i}} \right),
\end{align}
where $Z$ is the probability normalization factor, $k_j^+$ and $k_j^-$ are the binding and unbinding rates for ligand $j \in \{s,in \}$, respectively, and $\tau_{u,i}$ and $\tau_{b,i}$ are the $i^\text{th}$ observed unbound and bound time durations, respectively \cite{mora2015physical}. 


In the diffusion-limited case, i.e., where the reaction rates are much higher than the characteristic rate of diffusion, the binding rate can be simply written as $k^+ = 4 D a$ for circular receptors \cite{mora2015physical}, with $D$ and $a$ being the diffusion coefficient of ligands and the effective receptor size, respectively. Assuming that the ligands are of similar size, the diffusion coefficient $D$, which depends on the ligand size as well as the temperature and viscosity of the fluid medium \cite{bialek2012biophysics}, can be taken equal for all ligand types. In this case, the likelihood function \eqref{eq:likelihood1} reduces to
\begin{align}
\p\left(\{\tau_b,\tau_u \}_{N}\right) = \frac{1}{Z} e^{- T_u  k^+ c_{tot}}  (k^+ c_{tot})^N \prod_{i=1}^{N} \p\left(\tau_{b,i}\right) ,
\label{eq:likelihood}
\end{align}
where $T_u = \sum_{i = 1}^N \tau_{u,i}$ is the total unbound time of all receptors during the observation period, $c_{tot} = c_s + c_{in}$ is the total ligand concentration in the vicinity of the receptors, and $\p(\tau_{b,i})$ is the probability of observing a bound time duration, which is given as a \textit{mixture of exponential distributions}, i.e.,  
%
\begin{equation}
\p(\tau_b)=  \sum_{i \in \{s, in\}} \alpha_i k_i^- \e^{-k_i^- \tau_b} \label{eq:probboundduration}
\end{equation}
Here $\alpha_i = c_i/c_{tot}$ is the concentration ratio of a particular type of molecule.

The log-likelihood function for an observed set of unbound/bound time durations can then be written as the sum of three terms, i.e., 
\begin{align}\label{log_like}
\mathcal{L}(\{\tau_b,\tau_u \}_N) & = \ln\p(\{\tau_b,\tau_u \}_N) \\ \nonumber
& = \mathcal{L}_0  + \mathcal{L}\left(T_u | c_{tot}\right)  + \mathcal{L} \left(\{\tau_b \} | \bm{\alpha}\right), 
\end{align}
where $\mathcal{L}_0$ comprises the terms that do not depend on $c_{tot}$ or $\bm{\alpha} = [\alpha_{in}, \alpha_s]^T$, while $\mathcal{L}\left(T_u | c_{tot}\right) $ and $\mathcal{L} \left(\{\tau_b \} | \bm{\alpha}\right)$ are the functions of the total concentration $c_{tot}$ and the ligand concentration ratios $\bm{\alpha}$, respectively. For detection, we are only interested in the log-likelihoods that are functions of $c_{tot}$ and $\bm{\alpha}$, i.e.,
\begin{equation}
\mathcal{L}\left(T_u | c_{tot}\right) = N \ln(c_{tot})-k^+ c_{tot} T_u, \label{eq:likelihoodtotalligand}
\end{equation}
\begin{equation}\label{l2}
\mathcal{L} \left(\{\tau_b \} | \bm{\alpha}\right) = \sum_{i=1}^{N} \ln \p\left(\tau_{b,i}\right).
\end{equation}
Accordingly, $\mathcal{L}\left(T_u | c_{tot}\right)$ tells us that the total unbound time $T_u$ is informative of the total ligand concentration $c_{tot}$, whereas $\mathcal{L} \left(\{\tau_b \} | \bm{\alpha}\right)$ shows that the individual bound time durations $\{\tau_b \}$ are informative of the ligand concentration ratios $\bm{\alpha}$. 

\section{MC System}
\label{sec:system}
We consider an MC system with a receiver equipped with single type of ligand receptors, attempting to decode a binary message $s \in \{0,1 \}$ encoded by a distant transmitter into the concentration of molecules, i.e., $c_s$, which propagate in the liquid MC channel through free diffusion. The following assumptions are adopted to define the system:
\begin{itemize}
	\item Following the convention in MC literature \cite{pierobon2011noise}, receiver is assumed to have a reception space of a volume $V$ around its lipid membrane, in which receptors along with incoming information and interferer molecules are uniformly distributed at any time.
	
	\item Receiver is time synchronized with the transmitter. In the absence of interferer molecules, the receiver has perfect knowledge of the channel state information (CSI) such that it exactly knows the concentration of information molecules in the reception space corresponding to $s=0$ and $s=1$ transmissions, i.e., $c_{s=0}$ and $c_{s=1}$, respectively. This is justified by the fact that molecular concentration at any point in three dimensional free diffusion channel is deterministically governed by the Fick's second law of diffusion \cite{pierobon2011diffusion}. On the other hand, in the presence of interferer molecules, receiver only knows the probability distribution of the number of interferer molecules in the reception space.  Our analysis will not explicitly take the intersymbol interference (ISI) into account for tractability of the derivations; however, we will perform analyses in Section \ref{sec:performance} for cases when the receptors approach saturation as a result of high-level ISI.  
	
	\item Sampling is performed only once for each receptor in a single signaling interval, such that the number of samples taken for each transmission is equal to the number of receptors, i.e., $N = N_R$.  Received molecular signal is taken as steady around the sampling time assuming that the MC system manifests diffusion-limited characteristics, i.e., diffusion dynamics are much slower than the binding kinetics. Receptors are assumed to be operating independently of each other. 
	

	\item The channel and the reception space of the receiver are crowded with interferer molecules, which can bear significant affinity with the receptors. The number of interferer molecules $n_{in}$ in the reception space during a sampling period is taken as a Poisson random variable with the mean number $\mu_{n_{in}}$ \cite{bialek2012biophysics}. The binding rates of information and interferer molecules are taken as equal, i.e., $k_s^+ = k_{in}^+ = k^+$, following the assumption of diffusion-limited binding kinetics, as discussed in Section \ref{sec:statistics}.  However, the unbinding rates, determined by the affinity with the receptors, are different for information and interferer molecules, and denoted by $k_s^-$ and $k_{in}^-$, respectively. 
	
	
\end{itemize}

\section{Detection Methods}
\label{sec:detection}
We introduce four detection methods that use different observable statistics of ligand-receptor binding reaction to decode the incoming messages in the presence of a random number of interferer molecules. These detection methods are based on the instantaneous number of bound receptors, total unbound time of receptors, receptor bound time intervals, and the combination of total unbound time time of receptors and receptor bound time intervals. 



\subsection{Detection based on Number of Bound Receptors (DNBR)}

The simplest detection method, which is widely studied in the MC literature, is based on sampling the number of bound receptors, exploiting the relation between ligand concentration and binding probability. As reviewed in Section \ref{sec:statistics}, the probability of finding a receptor in the bound state in the presence of interferers is given as
\begin{equation} \label{probBinding}
\p(\B|s, n_{in})  = \frac{c_s/K_D^s + c_{in}/K_D^{in}  }{1 + c_s/K_D^s + c_{in}/K_D^{in}},
\end{equation}
where $c_{in} = n_{in}/V$. Note that we condition the probability on the number of interferer molecules, $n_{in}$, instead of their concentration for mathematical convenience in dealing with the discrete Poisson distribution. 

As we discussed in Section \ref{sec:statistics}, the probability distribution of the number of bound receptors is binomial. Hence, given the number of information and interferer molecules in the reception space, the mean and variance of number of bound receptors $n_\B$ at equilibrium can be written as follows
\begin{align}
\E[n_\B|s, n_{in}] &= \p(\B|s, n_{in})  N_R, \\ \nonumber
\Var[n_\B|s, n_{in}] &= \p(\B|s, n_{in})  \bigl(1-\p(\B|s, n_{in}) \bigr) N_R.  \label{eq:Binomial}
\end{align}

%
The mean and variance conditioned only on the number of information molecules thus can be obtained by applying the total law of expectation and variance, i.e., 
\begin{align}
\E[n_\B|s] &=   \sum_{n=0}^\infty \E[n_\B|s,n_{in} = n] \p(n_{in} = n),  \\ \nonumber
\Var[n_\B|s] &= \E\Bigl[\Var[n_\B|s, n_{in}]\Big|s \Bigr]  + \Var\Bigl[\E[n_\B|s, n_{in}]\Big|s \Bigr],  \label{eq:numberofboundMeanAndVariance}
\end{align}
which do not lend themselves into a more tractable form, and therefore, the summations should be performed until the results converge. It is shown in Fig. \ref{fig:Gaussian_n} that the resulting probability distribution can be well approximated with a Gaussian distribution. Hence, we can write 
\begin{equation}
\p(n_\B|s) \approx \mathcal{N}\Bigl(\E[n_\B|s] , \Var[n_\B|s] \Bigr). \label{eq:DNBRpdf} 
\end{equation}

\subsection{Detection based on Receptor Unbound Time Durations  (DRUT)}

\begin{figure*}[!t]
	\centering
	\subfigure[]{
		\includegraphics[width=7cm]{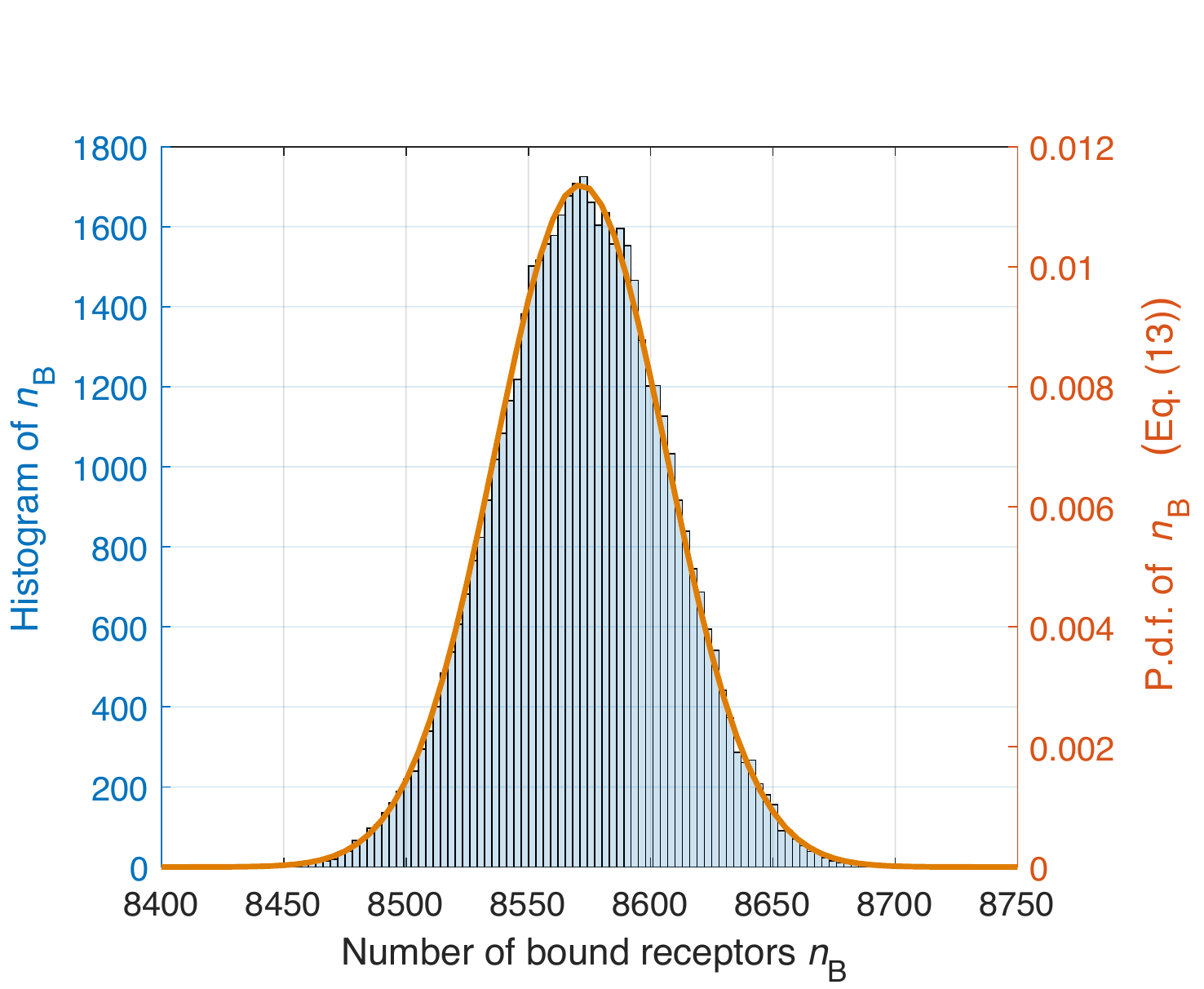}
		\label{fig:Gaussian_n}
	}
	\subfigure[]{
		\includegraphics[width=7.2cm]{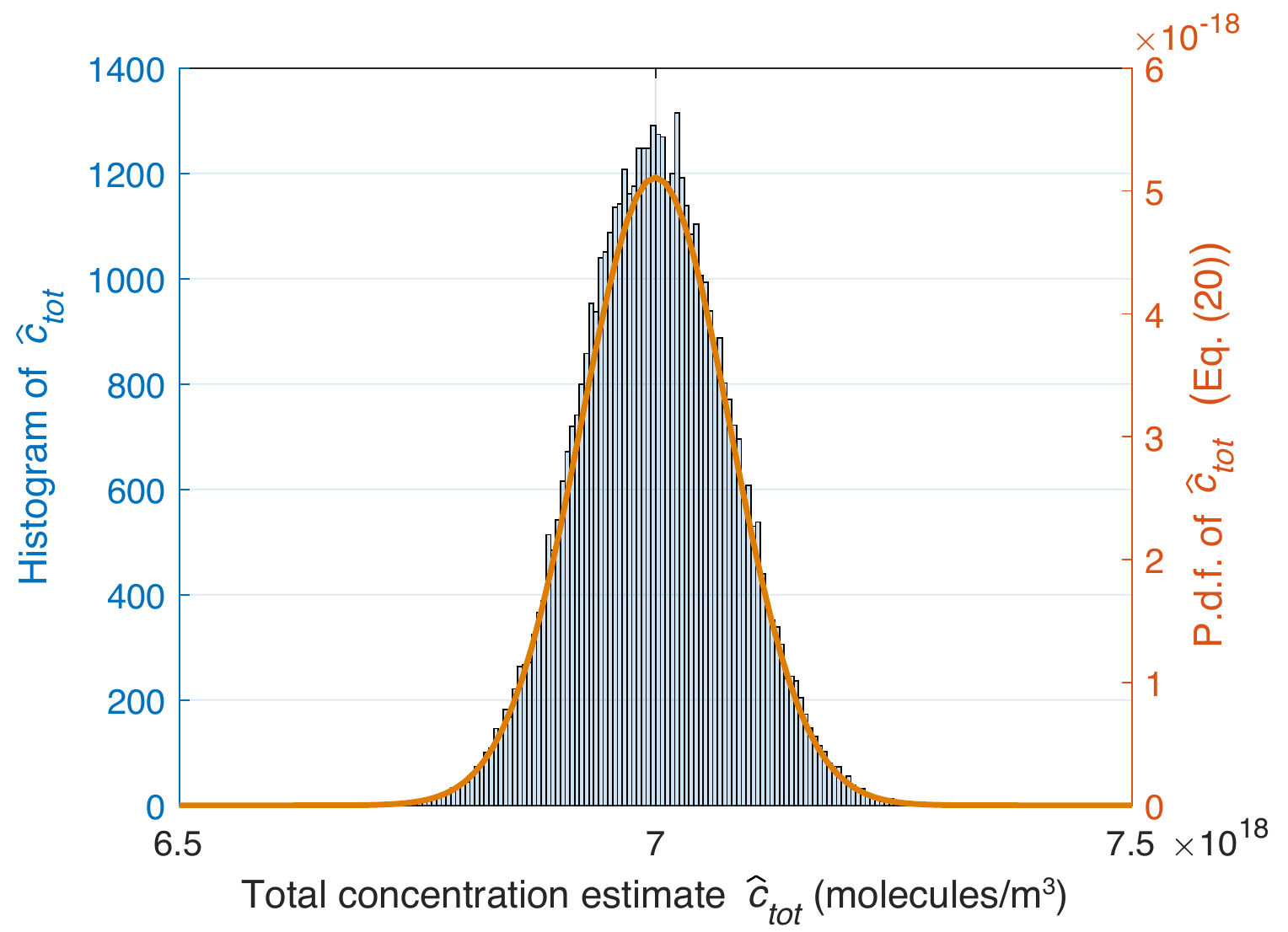}
		\label{fig:Gaussian_ctot}
	}
	\subfigure[]{
		\includegraphics[width=7cm]{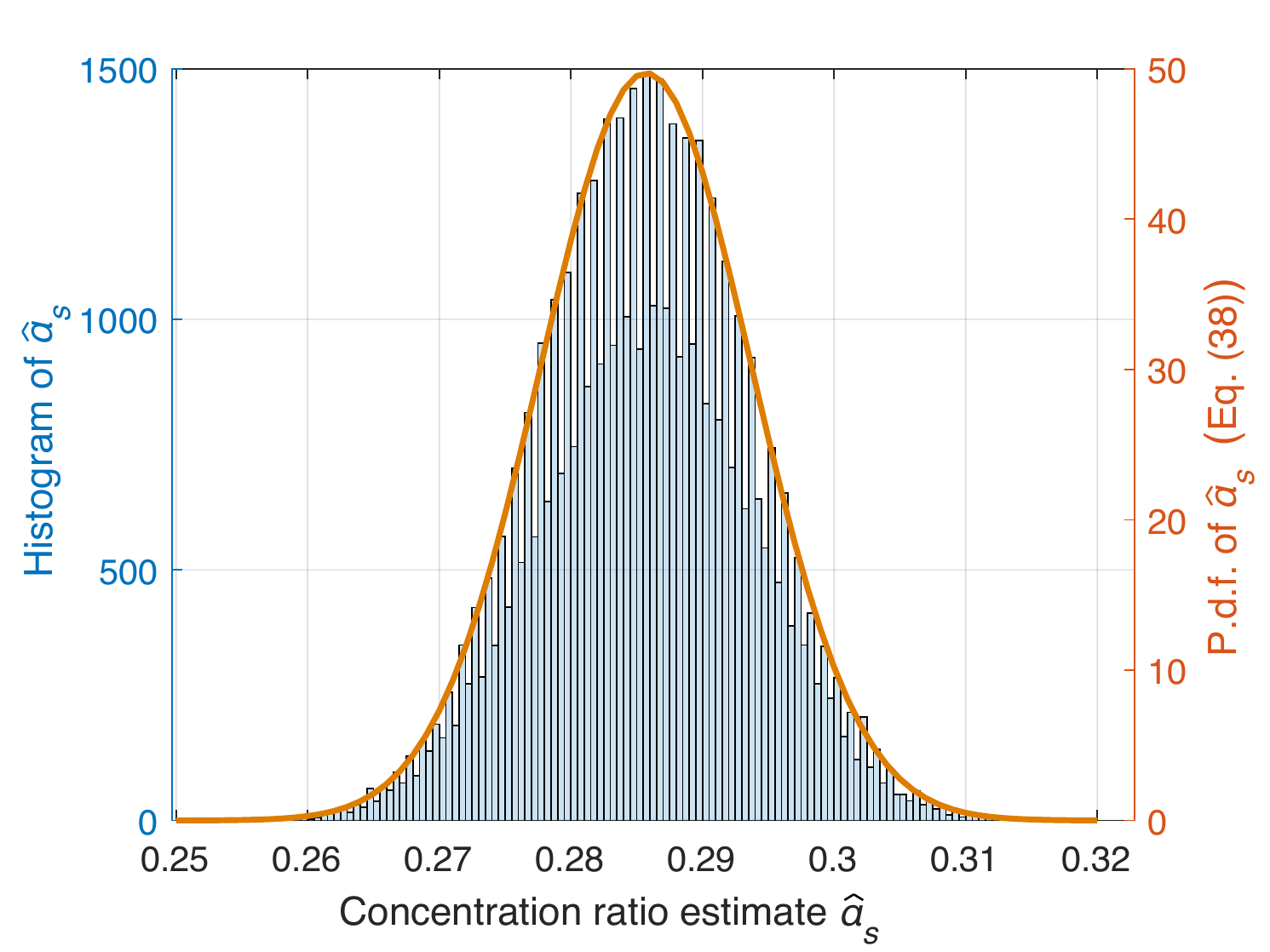}
		\label{fig:Gaussian_alpha}
	}
	\subfigure[]{
		\includegraphics[width=7cm]{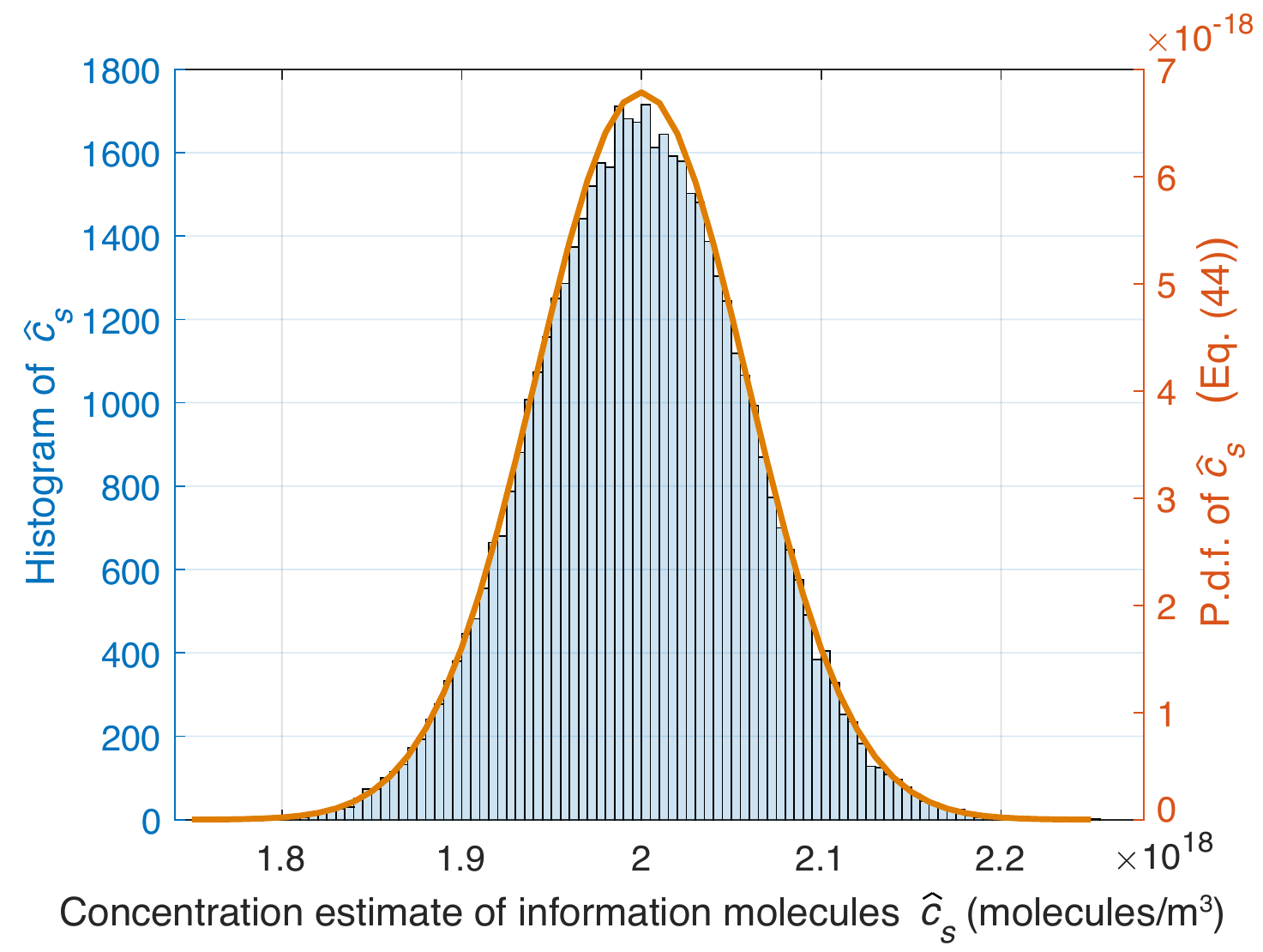}
		\label{fig:Gaussian_c}
	}
	\caption{Gaussian approximation of decision statistics. Histograms are obtained via Monte Carlo simulations (50000 iterations) of stochastic ligand-receptor binding process under interference. Simulation parameters are set to the default values that are used in performance evaluation (see Table \ref{table:parameters}). Here we assume bit-0 is transmitted. Hence, we use $c_0 = 4 \times K_{D,S}$.  }
	\label{fig:Gaussian}
\end{figure*}

In this method, the receiver performs the detection based on the estimation of total ligand concentration in the reception space using the total unbound time duration of receptors. From the log-likelihood function \eqref{eq:likelihoodtotalligand}, we can obtain the ML estimate of the total molecule concentration as follows: 
\begin{align}
\frac{\partial \mathcal{L}\left(T_u | c_{tot}\right)}{\partial c_{tot}} \bigg|_{\hat{c}_{tot}^\ast} = 0,
\end{align}
which yields the ML estimator $\hat{c}_{tot}^\ast = \frac{N}{k^+ T_U}$. However, this is a biased estimator unless $N$ is very large. An unbiased version of this estimator can be obtained with a slight modification \cite{kuscu2018maximum} as follows
\begin{align}
\hat{c}_{tot} =  \frac{N-1}{k^+ T_u} \label{ctot_estimator},
\end{align}
whose mean and variance are obtained as 
\begin{align}
\E[\hat{c}_{tot}|s, n_{in}] &=  c_{tot} = c_s + n_{in}/V \label{Tu_mean},\\
\Var[\hat{c}_{tot}|s, n_{in}] &= \frac{c_{tot}^2}{N - 2}  = \frac{(c_s + n_{in}/V )^2}{N -2}\label{Tu_variance}.
\end{align}
Using the law of total expectation and variance, the mean and variance of this estimator conditioned only on the number of information molecules can be written as 
\begin{align}
\E[\hat{c}_{tot}|s] &=  \E\Bigl[\E[\hat{c}_{tot}|s, n_{in}]\Big|s\Bigr] \\ \nonumber
&= \E\bigr[c_s + n_{in}/V|s\bigr] \\ \nonumber
&= c_s + \mu_{n_{in}}/V,
\end{align}
\begin{align}
&\Var[\hat{c}_{tot}|s] \\ \nonumber
&= \E\Bigl[\Var[\hat{c}_{tot}|s, n_{in}]\Big|s \Bigr] + \Var\Bigl[\E[\hat{c}_{tot}|s, n_{in}]\Big|s \Bigr]  \\ \nonumber
&= \mathlarger{\sum}_{n=0}^\infty \frac{(c_s + n/V)^2}{ (N-2)} \p(n_{in} = n) + \Var\left[c_s + n_{in}/V |s\right] \\ \nonumber
& \approx \frac{1}{(N-2) \sqrt{2 \pi \mu_{n_{in}}}}  \int_{-\infty}^\infty \Bigl(c_s + \frac{n}{V}\Bigr)^2 \e^{ -\frac{(n-\mu_{n_{in}})^2}{2\mu_{n_{in}}}} dn + \frac{\mu_{n_{in}}}{V^2}\\ \nonumber
&\approx \frac{(c_s + \mu_{n_{in}}/V)^2} {N-2} + \frac{\mu_{n_{in}} (N-1)} {V^2 (N-2)},
\label{eq:numberofboundMeanAndVariance}
\end{align}
where the discrete Poisson distribution is approximated as a continuous Gaussian distribution. As demonstrated in Fig. \ref{fig:Gaussian_ctot}, the probability distribution of the total concentration estimator can also be approximated with a Gaussian distribution, i.e.,
\begin{equation}
\p(\hat{c}_{tot}|s) \approx \mathcal{N}\Bigl(\E[\hat{c}_{tot}|s] , \Var[\hat{c}_{tot}|s] \Bigr). \label{eq:DRUTpdf}
\end{equation}
\begin{figure}[!t]
	\centering
	\includegraphics[width=8cm]{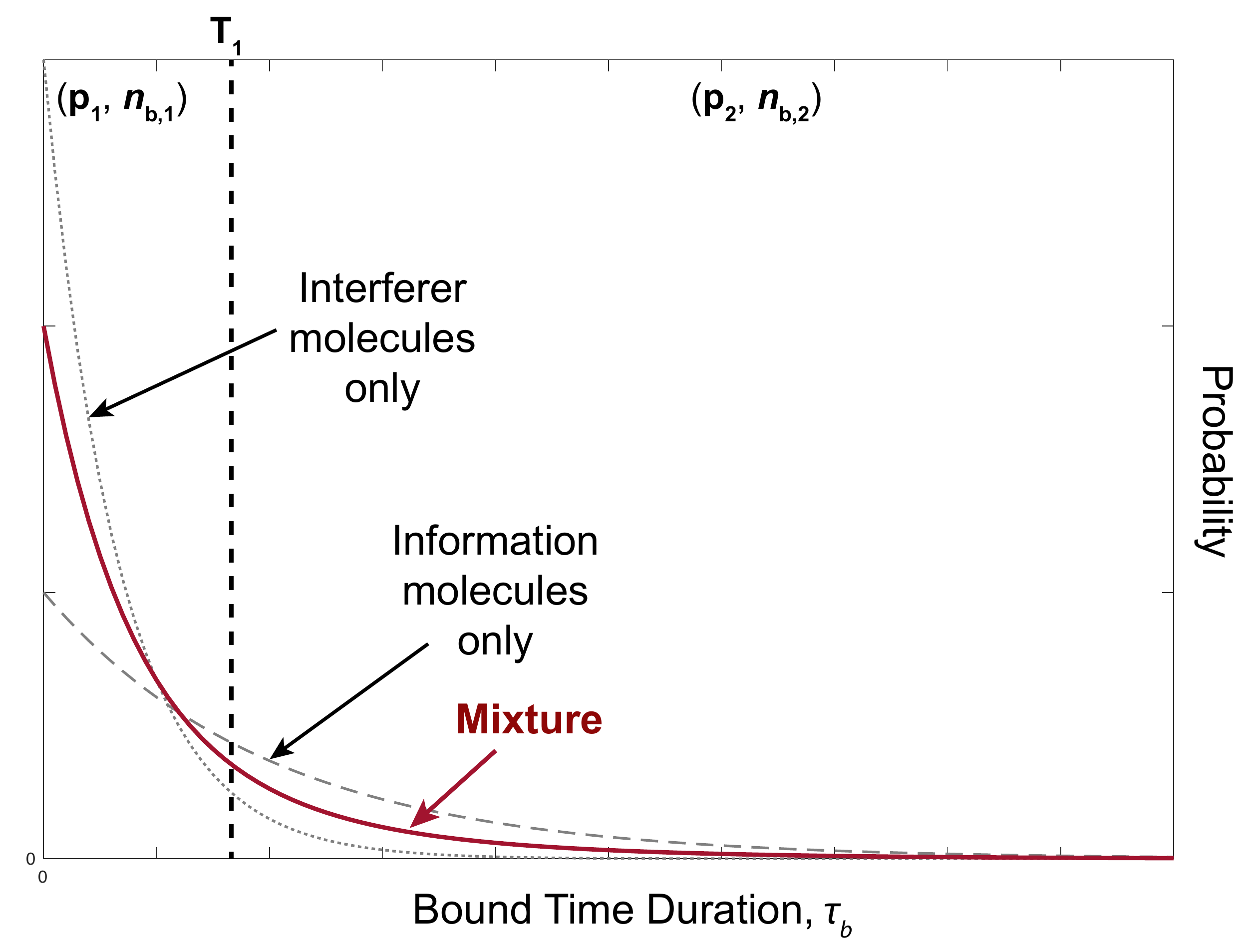}
	\caption{Probability distributions of the bound time durations corresponding to interferer molecules, information molecules, and the mixture of them. The dashed line indicates the time threshold (see \eqref{timethreshold}) that helps discriminate between information and interferer molecules by separating the longer binding events from the shorter ones. }
	\label{fig:mixture}
\end{figure}

\subsection{Detection based on Receptor Bound Time Durations  (DRBT)}
The concentration ratio of information molecules in the reception space, $\alpha_s$, is also expected to be different for bit-0 and bit-1 transmissions, and thus, can be used for detection. We can obtain the ML estimation for the concentration ratio of information molecules, i.e., $\hat{\alpha}_s^{ML}$, by solving $\partial \mathcal{L} \left(\{\tau_b \} | \bm{\alpha}\right)  / \partial \alpha_s \Big|_{\hat{\alpha}_s^{ML}} = 0$, i.e., 
\begin{equation}\label{l2}
0 =\mathlarger{\mathlarger{\sum}}_{i=1}^{N} \frac{k_s^- \e^{-\left(k_s^- \tau_{b,i}\right)}} {  \hat{\alpha}_s^{ML} k_s^- \e^{-k_s^- \tau_{b,i}} + (1-\hat{\alpha}_s^{ML}) k_{in}^- \e^{-k_{in}^- \tau_{b,i}}}.
\end{equation}
However, the expression in \eqref{l2} does not have any analytical solution for ML estimate $\hat{\alpha}_s^{ML}$, and requires numerical approaches, which are not feasible for resource-limited bionanomachines \cite{kuscu2019transmitter}. Instead, in \cite{kuscu2019channel}, we proposed a feasible near-optimal estimation method for concentration ratios based on counting the number of binding events that fall in specific time intervals. In this estimation scheme, the time domain is divided into as many time intervals as the number of molecule types existing in the channel. In the presence of information molecules and only one type of interferer molecules, we need two time intervals, as demonstrated in Fig. \ref{fig:mixture}. These non-overlapping intervals are demarcated by a time threshold, which can be taken as proportional to the inverse of the unbinding rate of the molecule type with the lower binding affinity. Given that the interferer molecules bind receptors with lower affinity, we can write the time threshold as 
\begin{equation} \label{timethreshold}
T_1 = \nu/k_{in}^-.
\end{equation}
Here, $\nu > 0$ is the proportionality constant. In this paper, we use  $\nu = 3$, which was previously shown in \cite{kuscu2019channel} to yield near-optimal performance in terms of estimation error. Note that in the same paper, we also showed that this transduction scheme is suitable for biological MC devices, as it can be implemented by active receptors based on the well-known KPR scheme. 

The probability of observing a ligand binding event with a binding duration falling in a specific time range can be obtained as

\begin{align}
\p_i = \int_{T_{i-1}}^{T_i} \p(\tau_b') d\tau_b' =&   \alpha_s \left( \e^{-k_s^- T_{i-1}} - \e^{-k_s^- T_{i}} \right) \\ \nonumber
&+ \alpha_{in} \left( \e^{-k_{in}^- T_{i-1}} - \e^{-k_{in}^- T_{i}} \right),
\end{align}
where $\p(\tau_b)$ is given in \eqref{eq:probboundduration}. Here we take $T_0 = 0$ and $T_2 = +\infty$. In matrix notation, the probabilities can be written as
\begin{equation} \label{eq:probmatrix}
\bm{\p} = \bm{Q} \bm{\alpha},
\end{equation}
where $\bm{\p} = [\p_1, \p_2]^T$,  $\bm{\alpha} = [\alpha_{in}, \alpha_s]^T$, and $\bm{Q}$ is an ($2 \times 2$) matrix with the elements 
%

\[
\bm{Q}=
\begin{bmatrix}
q_{1,1} & q_{1,2} \\
q_{2,1} & q_{2,2}
\end{bmatrix}
=
\begin{bmatrix}
1 - \e^{-k^-_{in} T_1} & 1- \e^{-k^-_s T_{1}} \\
\e^{-k^-_{in} T_{1}} & \e^{-k^-_s T_{1}}
\end{bmatrix}
\]

The number of binding events that fall in each time range follows a binomial distribution with the mean and the variance given by
\begin{equation}
\bm{\E[n_b}|s,n_{in}\bm{]} = \bm{\p} N,
\label{eq:meann}
\end{equation}
\begin{equation}
\bm{\Var[n_b}|s,n_{in}\bm{]} = \bigl( \bm{\p} \odot (1-\bm{\p}) \bigr) N,
\label{eq:varn}
\end{equation}
where $\bm{n_b}$ is an $(2 \times 1)$ vector with the vector elements $n_{b,i}$ being the number of binding events, whose durations are within the $i^\text{th}$ time range demarcated by $T_{i-1}$ and $T_i$, and $\odot$ denotes the Hadamard product, i.e., $(\bm{K} \odot \bm{L})_{i,j} = (\bm{K})_{i,j} (\bm{L})_{i,j}$. 

Applying the Method of Moments (MoM) with only the first moment yields a concentration ratio estimator in terms of number of binding events \cite{kuscu2019channel}, i.e., 
\begin{align} \label{Wmatrix}
\bm{\hat{\alpha}} = \left(\frac{1}{N}\right) \bm{W} \bm{n_b},  
\end{align}
where $\bm{W} = \bm{Q}^{-1}$, i.e., the inverse of $\bm{Q}$ matrix, which is also a ($2 \times 2$) matrix with elements $w_{i,j}$. 
Note that the estimated concentration ratio of information molecules becomes
\begin{align}
\hat{\alpha}_s =  \left( n_{b,1} w_{2,1} + n_{b,2} w_{2,2} \right)/N. \label{alpha_estimator}
\end{align}


The variance of this ratio estimator can be written as
\begin{align}\label{eq:alpha_variance}
\Var[&\hat{\alpha}_s|s, n_{in}] \\ \nonumber
&= \frac{1}{N^2}  \sum_{i=1}^2 \sum_{j=1}^2 w_{2,i} w_{2,j} ~\Cov[n_{b,i}, n_{b,j}|s, n_{in}],  
\end{align}
with the covariance function
\begin{equation}
\Cov[n_{b,i}, n_{b,j}| s, n_{in}]  = 
\begin{cases}
\Var[n_{b,i}| s, n_{in}],  & \text{if } i = j,\\
-p_i p_j N,    & \text{otherwise}.
\end{cases}
\label{eq:est_covariance}
\end{equation}
After some trivial mathematical manipulations, we can rewrite \eqref{eq:alpha_variance} in closed form as 
\begin{align}
\Var[\hat{\alpha}_s|s, n_{in}] = \frac{1}{N} \frac{\Gamma_1 (n_{in}/V)^2 + \Gamma_2 (n_{in}/V) + \Gamma_3}{(c_s + n_{in}/V)^2},  \label{eq:est_variance}
\end{align}
where, $\Gamma_1$, $\Gamma_2$ and $\Gamma_3$ are given in \eqref{eq:Gamma1}, \eqref{eq:Gamma2}, and \eqref{eq:Gamma3}, respectively.
\addtocounter{equation}{3}  

The expected value of the ratio estimator is equal to the actual value of the concentration ratio vector $\bm{\alpha}$, i.e.,
\begin{align}
\bm{\E[\hat{\alpha}}|s, n_{in} \bm{]} &= \frac{1}{N} \bm{W} \bm{\E[n_b}|s,n_{in}\bm{]}  \\ \nonumber
&= \bm{W} \bm{\p} = \bm{Q^{-1}} \bm{\p} = \bm{\alpha}.
\label{eq:meanratioest}
\end{align}

Using the law of total expectation and variance, we can write
\begin{align}
\E[\hat{\alpha}_s|s] &=  \E\Bigl[\E[\hat{\alpha}_s|S, n_{in}]\Big|s\Bigr] \\ \nonumber
&= \E\biggl[\frac{c_s}{c_s + n_{in}/V} \bigg|s\biggr] \\ \nonumber
&= \mathlarger{\sum}_{n=0}^\infty \frac{c_s}{c_s + n/V} \p(n_{in} = n),
\end{align}
\begin{align}\label{eq:boundtimeMeanAndVariance}
&\Var[\hat{\alpha}_s|s] \\ \nonumber
&= \E\Bigl[\Var[\hat{\alpha}_s|s, n_{in}] \Big|s \Bigr] + \Var\Bigl[\E[\hat{\alpha}_s|s, n_{in}] \Big|s\Bigr] \\ \nonumber
&=  \mathlarger{\sum}_{n=0}^\infty \frac{1}{N} \frac{\Gamma_1 (n/V)^2 + \Gamma_2 (n/V) + \Gamma_3 + N c_s^2}{(c_s + n/V)^2} ~\p(n_{in} = n) \\ \nonumber
& ~~~- \left(\sum_{n=0}^\infty \frac{c_s}{c_s + n/V}  \p(n_{in} = n)\right)^2.
\end{align}
It is shown in Fig. \ref{fig:Gaussian_alpha} that the distribution of the ratio estimator for information molecules can be approximated with a Gaussian distribution as follows
\begin{equation}
\p(\hat{\alpha}_s|s) \approx \mathcal{N}\Bigl(\E[\hat{\alpha}_s|s] , \Var[\hat{\alpha}_s|s] \Bigr). \label{eq:DRBTpdf}
\end{equation}
\begin{figure*}[!t]
	\normalsize
	\setcounter{mytempeqncnt}{\value{equation}}
	\setcounter{equation}{31}
	\begin{align} \label{eq:Gamma1}
	\Gamma_1  = w_{2,1}^2 q_{1,1} - w_{2,1}^2 q_{1,1}^2 - 2 w_{2,1}w_{2,2}q_{1,1}q_{2,1} + w_{2,2}^2 q_{2,1} - w_{2,2}^2 q_{2,1}^2 
	\end{align}
	\begin{align} \label{eq:Gamma2}
	\Gamma_2  = & c_s  \bigl( w_{2,1}^2 q_{1,2} + w_{2,1}^2 q_{1,1}  - w_{2,1}^2 q_{1,1} q_{1,2} - w_{2,1}^2 q_{1,1} q_{1,2}  - 2 w_{2,1} w_{2,2} q_{1,1} q_{2,2}  - 2 w_{2,1} w_{2,2} q_{1,2} q_{2,1}  \\ \nonumber
	&+ w_{2,2}^2 q_{2,2}  + w_{2,2}^2 q_{2,1}   + w_{2,2}^2 q_{2,1} q_{2,2}  - w_{2,2}^2 q_{2,1} q_{2,2}\bigr)  
	\end{align}
	\begin{align} \label{eq:Gamma3}
	\Gamma_3  = c_s^2 \bigl(w_{2,1}^2 q_{1,2} - w_{2,1}^2 q_{1,2}^2 - 2 w_{2,1} w_{2,2} q_{1,2} q_{2,2} + w_{2,2}^2 q_{2,2} - w_{2,2}^2 q_{2,2}^2 \bigr) 
	\end{align}
	\setcounter{equation}{\value{mytempeqncnt}}
	\hrulefill
	\vspace*{4pt}
\end{figure*}

\subsection{Detection based on Receptor Unbound and Bound Time Durations (DRUBT)}
Combining the ratio estimator with the unbiased estimator of total ligand concentration, we can obtain an estimator for the individual concentration of information molecules as follows 
\begin{align} \label{eq:csestimator}
\hat{c}_s &= \hat{c}_{tot} \times \hat{\alpha}_s \\ \nonumber
&= \frac{N-1}{N} \frac{1}{k^+ T_u}  \left( n_{b,1} w_{2,1} + n_{b,2} w_{2,2} \right) \\ \nonumber
&\approx \frac{1}{k^+ T_u}  \left( n_{b,1} w_{2,1} + n_{b,2} w_{2,2} \right), ~~~ \text{for}~ N \gg 1.
\end{align}
The mean of this concentration estimator conditioned on the number of information and interferer molecules can be calculated as follows
\begin{align}
\E&[\hat{c}_s|s, n_{in}] = \E[\hat{c}_{tot}|s, n_{in}] \E[\hat{\alpha}_s|s, n_{in} ] = c_{tot} \alpha_s = c_s,
\end{align}
by exploiting the conditional independence of $\hat{c}_{tot}$ and $\hat{\alpha}_S$. The variance of this estimator can be obtained as follows
\begin{align} \label{eq:totalvarianceunbiased}
\Var[\hat{c}_s|s, n_{in}] &=\Var[\hat{c}_{tot}|s, n_{in}] \Var[\hat{\alpha}_s|s, n_{in}] \\ \nonumber
&+ \Var[\hat{c}_{tot}|s, n_{in}] \E[\hat{\alpha}_s|s, n_{in}]^2  \\ \nonumber
&+ \Var[\hat{\alpha}_s|s, n_{in}] \E[\hat{c}_{tot}|s, n_{in}]^2 \\ \nonumber
& \approx \frac{1}{N} \Bigl(\Gamma_1 (n_{in}/V)^2 + \Gamma_2 (n_{in}/V) + \Gamma_3 + c_s^2\Bigr) \\ \nonumber
& \text{~~~~for~} N \gg 1.  
\end{align}

Employing the law of total expectation and variance, we then obtain
\begin{align}
\E[\hat{c}_s|s] = \E\Bigl[\E[\hat{c}_s|s, n_{in}]\Big|s\Bigr] = \E[c_s|s] = c_s,
\end{align}
\begin{align} \label{eq:concentrationEstimatiorVariance}
&\Var[\hat{c}_s|s] = \E\Bigl[\Var[\hat{c}_s|s, n_{in}]\Big|s\Bigr] + \Var\Bigl[\E[\hat{c}_s|s, n_{in}]\Big|s\Bigr] \\ \nonumber
&= \frac{1}{N} \sum_{n=0}^\infty \Bigl(\Gamma_1 (n/V)^2 + \Gamma_2 (n/V) + \Gamma_3 + c_s^2 \Bigr) \p(n_{in} = n) \\ \nonumber
&\approx  \frac{1}{N \sqrt{2 \pi \mu_{n_{in}}}}  \\ \nonumber
&\times \int_{-\infty}^\infty \left(\Gamma_1 \left(\frac{n}{V}\right)^2+ \Gamma_2 \left(\frac{n}{V}\right) + \Gamma_3 + c_s^2\right) \e^{ -\frac{(n-\mu_{n_{in}})^2}{2\mu_{n_{in}}}} dn\\ \nonumber
&\approx \frac{1}{N} \left(\Gamma_1 \left(\frac{\mu_{n_{in}}}{V}\right)^2 + (\Gamma_1 + \Gamma_2) \frac{\mu_{n_{in}}}{V} + \Gamma_3 + c_s^2 \right).
\end{align}
Note that in \eqref{eq:concentrationEstimatiorVariance} we approximate the Poisson distribution of number of interferer molecules with a Gaussian distribution. 

We demonstrate in Fig. \ref{fig:Gaussian_c} that the p.d.f. of the concentration estimator can also be approximated with a Gaussian distribution, i.e.,
\begin{equation}
\p(\hat{c}_s|s) \approx \mathcal{N}\Bigl(\E[\hat{c}_s|s] , \Var[\hat{c}_s|s] \Bigr). \label{eq:DRUBTpdf} 
\end{equation}

\begin{table*}[!t]\scriptsize
	\centering
	\begin{threeparttable}
		\centering
		\caption[Comparison of detection methods under interference]{Comparison of Detection Methods}
		\label{table:comparison_interference}
		\begin{tabular}{llllll}
			\toprule	
			\textbf{Detection}   & \textbf{Sampled Receptor} & \textbf{Decision} & \textbf{Complexity}&  \textbf{Probability} \\ 
			\textbf{Method}   & \textbf{Statistics} & \textbf{Statistics} & &  \textbf{Distribution} \\
			\toprule
			DNBR  & Number of bound receptors & Number of bound receptors   & Low &  \eqref{eq:DNBRpdf}  \\ \midrule
			DRUT   & Total receptor unbound time & Total molecular concentration & Moderate & \eqref{eq:DRUTpdf} \\ \midrule
			DRBT         & Number of binding events of  & Concentration ratio of& High    &  \eqref{eq:DRBTpdf}    \\ 
			& durations in specific time intervals      & information molecules  &    &     \\ \midrule
			DRUBT &  Total receptor unbound time   & Total molecular concentration       & Very high &    \eqref{eq:DRUBTpdf} \\
			& + Number of binding events of   &  + Concentration ratio of& & \\
			&durations in specific time intervals    & information molecules & & \\  
			\bottomrule
		\end{tabular}%
	\end{threeparttable}	
\end{table*}%

\subsection{Decision Rule} 
\label{sec:detectionrule}
In previous sections, we obtain the likelihood of four different statistics given the number of information molecules in the reception space. A summary comparison of these detection methods is provided in Table \ref{table:comparison_interference}.

Considering that the system employs binary CSK, the decision rule can be simply written as
\begin{equation}
\hat{s} = \argmax_{s \in \{0,1\}} \p(\kappa|s),
\label{gaussian_DRBT2}
\end{equation}
where $\kappa \in \{n_\B, \hat{c}_{tot}, \hat{\alpha}_s, \hat{c}_s \}$ is the received signal statistics corresponding to the introduced detection methods. The decision rule can be further simplified by defining a decision threshold $\lambda_{\kappa}$, i.e.,
\begin{equation} 
\kappa \underset{{H_0}}{\overset{H_{1}}{\gtrless}} \lambda_{\kappa}.
\label{threshold}
\end{equation}
For normally distributed statistics, the optimal decision threshold yielding the minimum error probability can be calculated as follows
\begin{align} \label{threshold3}
& \lambda_{\kappa}  = \gamma_{\kappa}^{-1} \\ \nonumber
&\times  \Biggl( \Var[\kappa|s = 1] \E[\kappa|s = 0] -  \Var[\kappa|s = 0] \E[\kappa|s = 1]  \\ \nonumber
& + \Std[\kappa|s = 1]\Std[\kappa|s = 0] \\ \nonumber
& \times \sqrt{\bigl(\E[\kappa|s = 1] - \E[\kappa|s = 0] \bigr)^2+2 \gamma_{\kappa}  \ln{\frac{\Std[\kappa|s = 1]}{\Std[\kappa|s = 0]}}} ~\Biggr), 
\end{align}
where $\gamma_{\kappa} = \Var[\kappa|s = 1] - \Var[\kappa|s = 0]$, and $\Std[.] = \sqrt{\Var[.]}$ denotes standard deviation. Given the decision thresholds, the BEP for each detection method can be obtained as

\begin{align} \label{eq:bepbep}
\p_\kappa(e) &=\frac{1}{2} \bigg[ \p_\kappa(\hat{s} = 1|s=0) + \p_\kappa(\hat{s} = 0|s=1) \bigg] \\ \nonumber
&=\frac{1}{4}\Biggl[\erfc \Biggl(\frac{\lambda_\kappa - \E[\kappa|s = 0] }{\sqrt{2 \Var[\kappa|s = 0] }}\Biggr) \\ \nonumber 
&~~~+  \erfc \Biggl(\frac {\E[\kappa|s = 1]  - \lambda_\kappa}{\sqrt{2 \Var[\kappa|s = 1]} }\Biggr)\Biggr]. 
\end{align}

\section{Performance Evaluation}
\label{sec:performance}
In this section, we evaluate the performance of the introduced detection methods in terms of BEP, which is calculated according to \eqref{eq:bepbep}. The default values of the system parameters used in the analyses are given in Table \ref{table:parameters}, with the reaction rates adopted from the previous literature \cite{pierobon2011noise, kuscu2018modeling, bialek2012biophysics}.

Throughout the analysis, the amount of interferer molecules in the reception space is expressed in terms of their concentration. However, the following convention is adopted to convert the concentration into the number of molecules when dealing with the discrete Poisson distribution in the computations. 
\begin{align}
\mu_{n_{in}} &= \floor*{\mu_{c_{in}} V}.
\end{align}

We perform several analyses to evaluate the effect of the expected interferer concentration in the reception space, the ratio between the affinities of information and interferer molecules with the receptors, the ratio between the received information molecule concentrations corresponding to bit-$0$ ($s=0$) and bit-$1$ ($s=1$) transmissions,  and the number of receptors. In the presentation of the results, we also provide the saturation level of the receiver in terms of bound state probability $\p_\B$ of the receptors for $s=0$ and $s=1$.


\begingroup
\begin{table}[!t]
	\centering
	\caption{Default Values of System Parameters}
	\renewcommand{\arraystretch}{1.2} 
	\begin{tabular}{ l | l }
		\hline \hline
		Binding rate for both types of molecules $(k^+)$ & $2 \times 10^{-17}$ m$^3$/s \\ \hline
		Unbinding rate for information molecules $(k^-_s)$ &$  10 $ s$^{-1}$  \\ \hline
		Affinity ratio $(\eta)$ & $ 0.2  $\\ \hline
		Conc. of information molecules for $s = 0$ $(c_{s=0})$ & $4 \times K_D^s$  \\ \hline
		Conc. of information molecules for $s = 1$ $(c_{s=1})$ & $5 \times K_D^s$  \\ \hline
		Mean concentration of interferer molecules $(\mu_{c_{in}})$ & $2 \times K_D^{in}$ \\ \hline
		Number of receptors on the receiver surface $(N_R)$ & $10000$ \\ \hline
		Volume of the reception space $(V)$ & $4000~\mu$m$^3$ \\ \hline
	\end{tabular}
	\label{table:parameters}
\end{table}
\endgroup


\subsection{Effect of Interferer Concentration}
\begin{figure*}[!t]
	\centering
	\subfigure[]{
		\includegraphics[width=7.5cm]{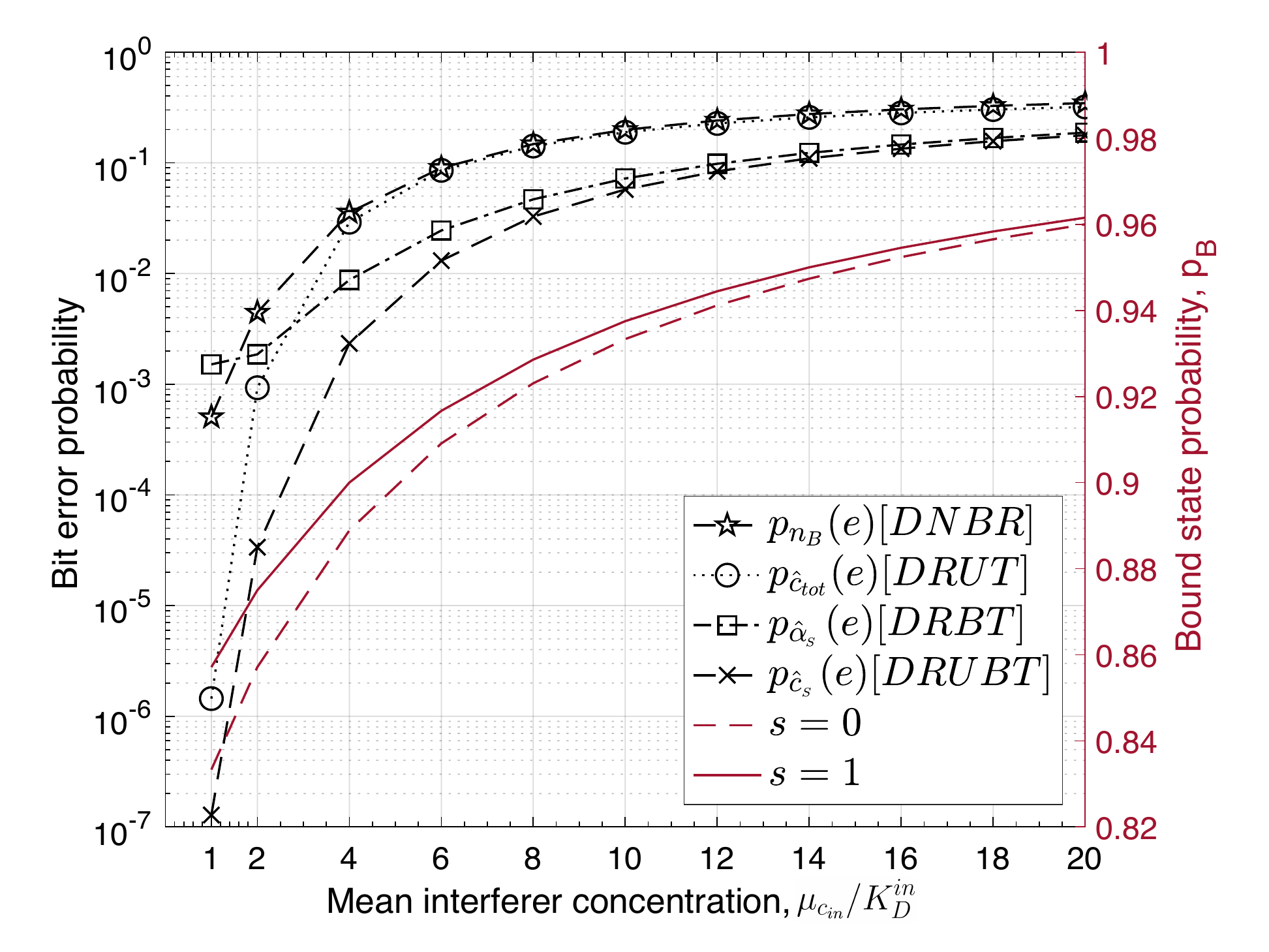}
		\label{fig:InterfererConcentration_NonSaturation}
	}
	\subfigure[]{
		\includegraphics[width=7.5cm]{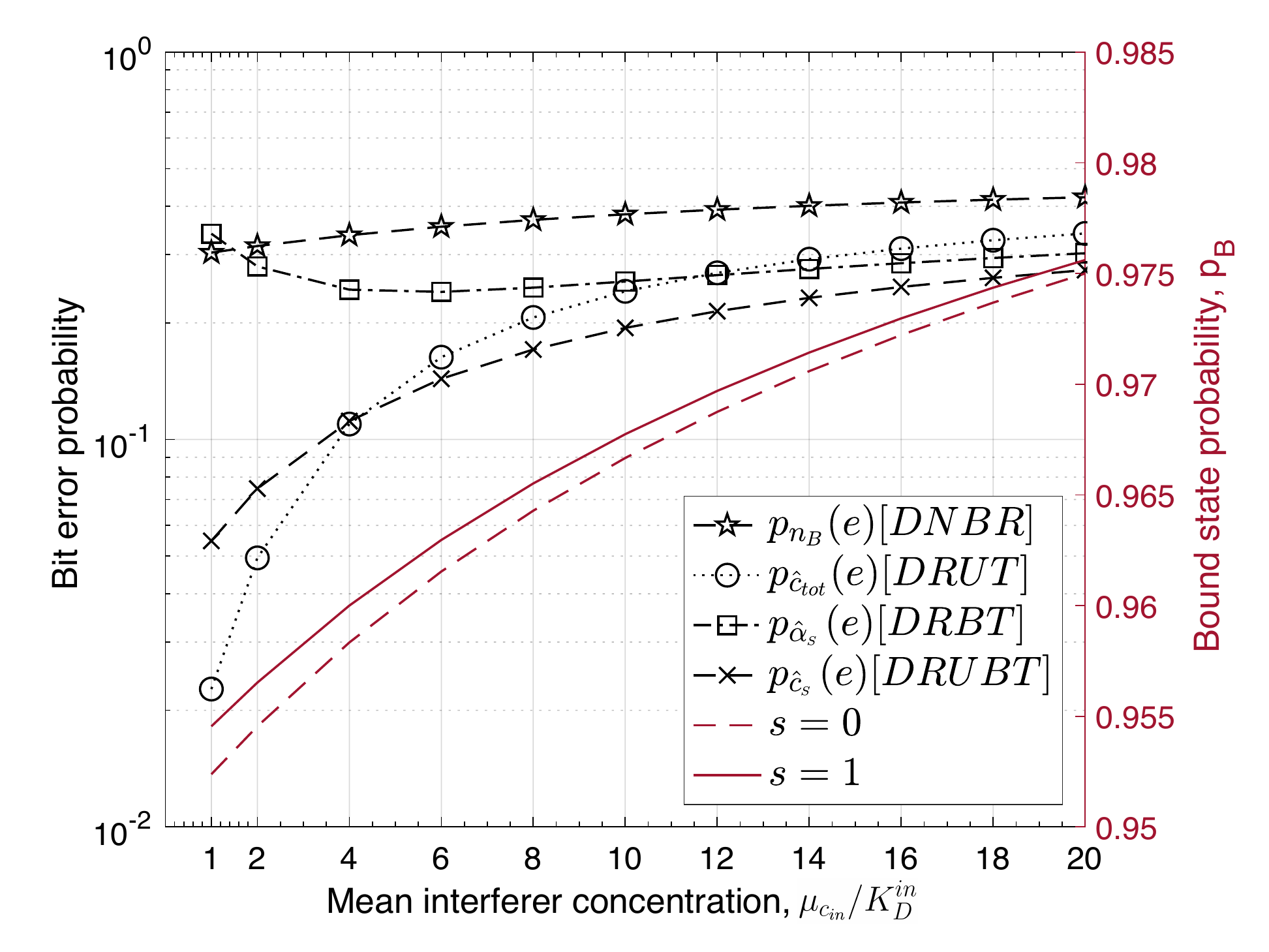}
		\label{fig:InterfererConcentration_Saturation}
	}
	\caption{Bit error probability as a function of mean interferer concentration $\mu_{c_{in}}/K_D^{in}$ for (a) non-saturation and (b) saturation conditions of the receiver.}
	\label{fig:InterfererConcentrationMain}
\end{figure*}

The first analysis concerns the strength of molecular interference. We analyze the effect of expected concentration of interferer molecules in the reception space for two scenarios differing in the receiver saturation level. In the first scenario, we consider that the receiver is reasonably away from saturation by setting the received concentration of information molecules for bit-0 and bit-1 as $c_{s=0} = 4K_D^s$ and $c_{s=1} = 5K_D^s$, respectively. The results, demonstrated in Fig. \ref{fig:InterfererConcentration_NonSaturation}, show that DRUBT outperforms the other detection methods in the simulated range of interference levels. On the other hand, DRUT, while performing poorly under high-level interference, substantially outperforms DNBR and DRBT when the interference level is relatively low. In the same region, the performance improvement obtained by DRUBT is more pronounced. It is also worth noting that DRNB, which is the simplest among the investigated detection methods, performs better than DRBT at the lowest level of interference. 

In the second case, the receptors are driven into saturation by setting the received concentrations as $c_{s=0} = 19K_D^s$ and $c_{s=1} = 20K_D^s$. In these conditions, the BEP for all detection methods is significantly higher than the non-saturation case as shown in Fig. \ref{fig:InterfererConcentration_Saturation}. Yet the performance improvement obtained with DRUBT is notable. This time, however, at low interference levels, DRUT, which was previously proposed for overcoming the receiver saturation problem \cite{kuscu2018maximum}, outperforms DRUBT.  
Also, in the case of receiver saturation, the performance of DRBT is worsened as the mean interference level decreases below $\mu_{c_{in}} = 4 K_D^{in}$, in contrast to the common trend observed in other detection methods. This is because the difference between the mean of ratio estimates conditioned on bit-0 and bit-1, i.e., $\E[\hat{\alpha}_s|s=0]$ and $\E[\hat{\alpha}_s|s=1]$, is a concave function of mean interferer concentration, which is maximized around $\mu_{c_{in}} = 4 K_D^{in}$. On the other hand, the corresponding variances of the ratio estimates monotonically increase with the decreasing interference level. This hampers the receiver's capacity to discriminate between bit-$0$ and bit-$1$. The poor performance of the ratio estimation in this range also affects the performance of DRUBT. As a result, at very low interference levels, DRUBT is outperformed by DRUT.



\subsection{Effect of Similarity between Information and Interferer Molecules}
The affinity ratio $\eta = k_s^-/k_{in}^-$ between information and interferer molecules determines how similar they are in terms of binding affinity with the receptors. The effect of affinity ratio on the detection performance is analyzed in two parts corresponding to the scenarios when $\eta<1$ and $\eta>1$. In both analyses, we keep the unbinding rate of information molecules constant and equal to its default value $k_s^- = 10$ s$^{-1}$, and thus, the unbinding rate of interferer molecules $k_{in}^-$ changes with the varying affinity ratio. We also keep the mean concentration of interferer molecules in the reception space constant and equal to 
$\mu_{c_{in}} = 10 K_D^s = 5$ $\mu$m$^{-3}$.
\begin{figure*}[!t]
	\centering
	\subfigure[]{
		\includegraphics[width=7.5cm]{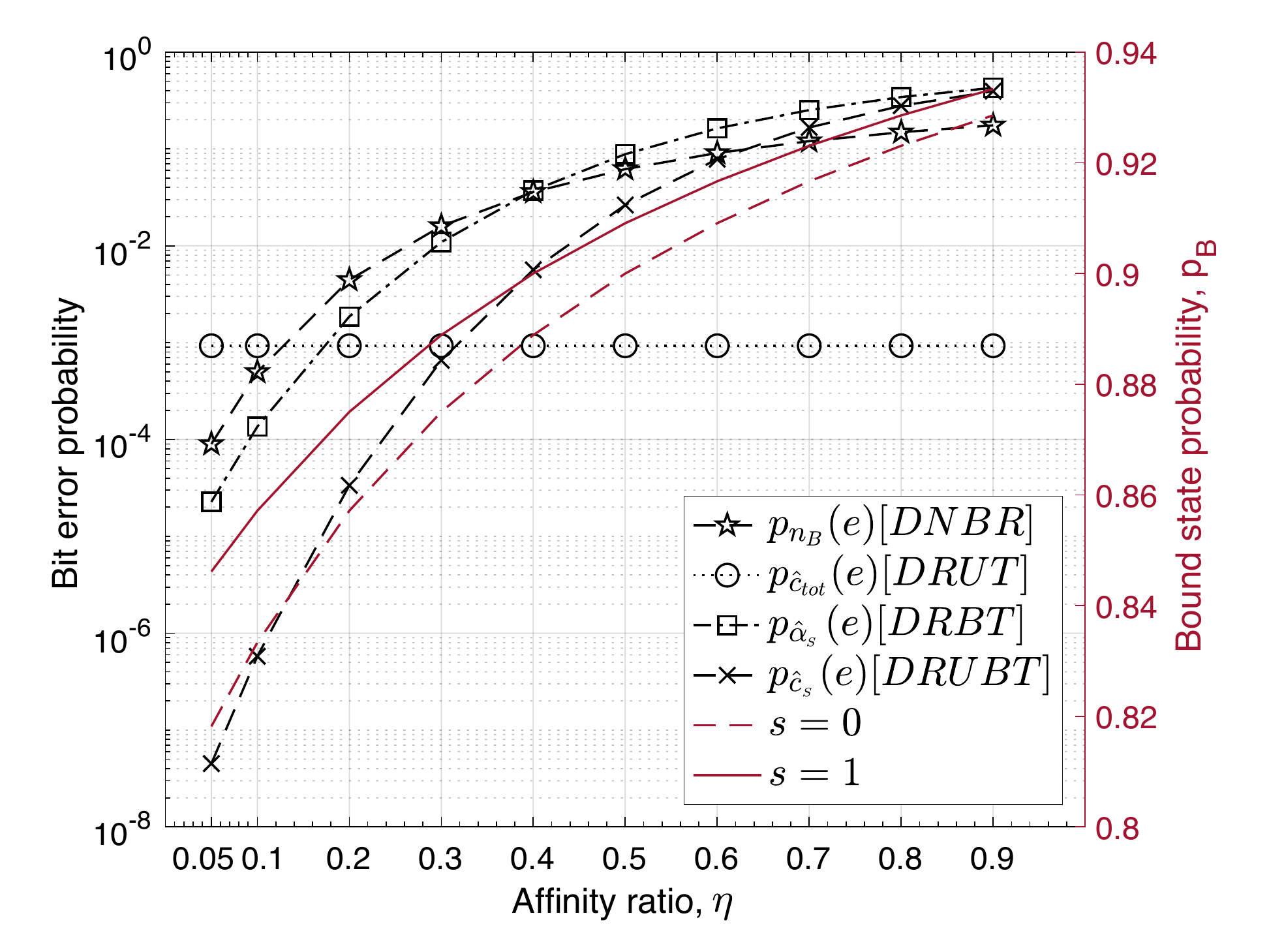}
		\label{fig:SimilarityAlpha}
	}
	\subfigure[]{
		\includegraphics[width=7.5cm]{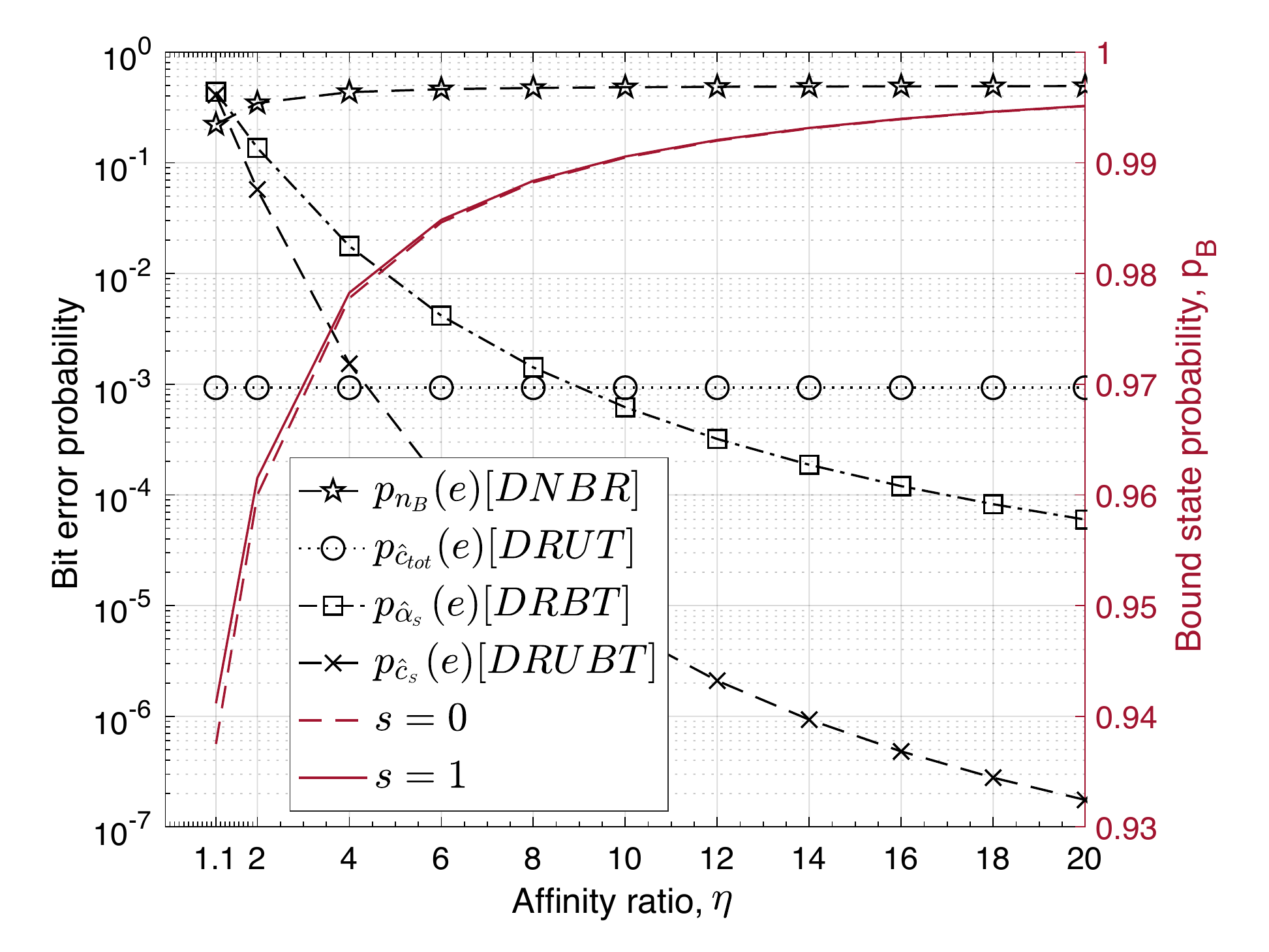}
		\label{fig:SimilarityAlphaReverse}
	}
	\caption{Bit error probability as a function of affinity ratio $\eta = k_s^-/k_{in}^-$ for the cases (a) when information molecules have more binding affinity, i.e., $\eta <1$, and (b) when interferer molecules have more binding affinity, i.e., $\eta >1$. }
	\label{fig:SimilarityAlphaMain}
\end{figure*}

In the first scenario, information molecules have higher binding affinity than the interferers, e.g., $\eta<1$. As is seen from the results provided in Fig. \ref{fig:SimilarityAlpha}, when the two types of molecules become more similar, the error probability substantially increases for all detection methods except DRUT. The performance of DRUT is not affected by the binding affinity, as the total unbound time of receptors, which DRUT solely relies on, is independent of the unbinding rate of bound ligand-receptor pairs, and only depends on the binding rate and the molecular concentration. On the other hand, DRUBT performs the best among the analyzed detection methods, when the information and interferer molecules differ greatly in terms of binding affinity. In the case of high similarity, the ratio estimator DRBT is the worst performer, as it becomes unable to discriminate between the two types of molecules based on their bound time durations. The same reasoning applies to the performance of DRUBT in this region, which also partly relies on the estimation of the concentration ratio. 

In the second analysis, we consider the case when the information molecules have lower binding affinity than the interferers. In this case, we expect that the interferer molecules occupy more receptors than the information molecules. As shown in Fig. \ref{fig:SimilarityAlphaReverse}, DNBR, the simplest detection method, performs particularly poorly for $\eta>1$.  As the affinity ratio increases, the advantage of the detection methods relying on the difference between unbinding rates becomes more pronounced. While DRUBT significantly outperforms other methods in this particular case, the performance of DRUT, which is much more practical than DRUBT, is notable. 

\begin{figure*}[!t]
	\centering
	\subfigure[]{
		\includegraphics[width=7.5cm]{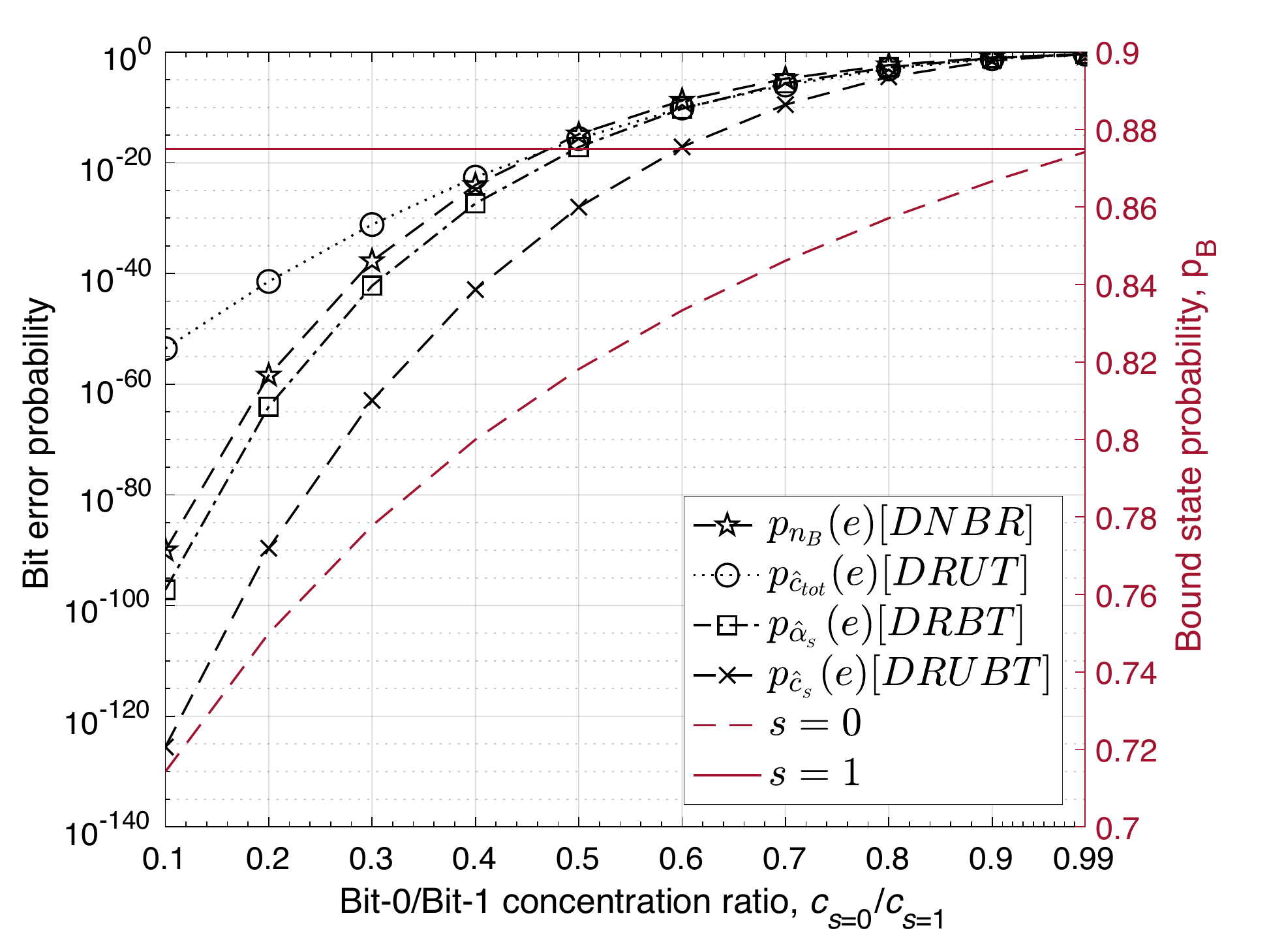}
		\label{fig:Bit0Concentration}
	}
	\subfigure[]{
		\includegraphics[width=7.5cm]{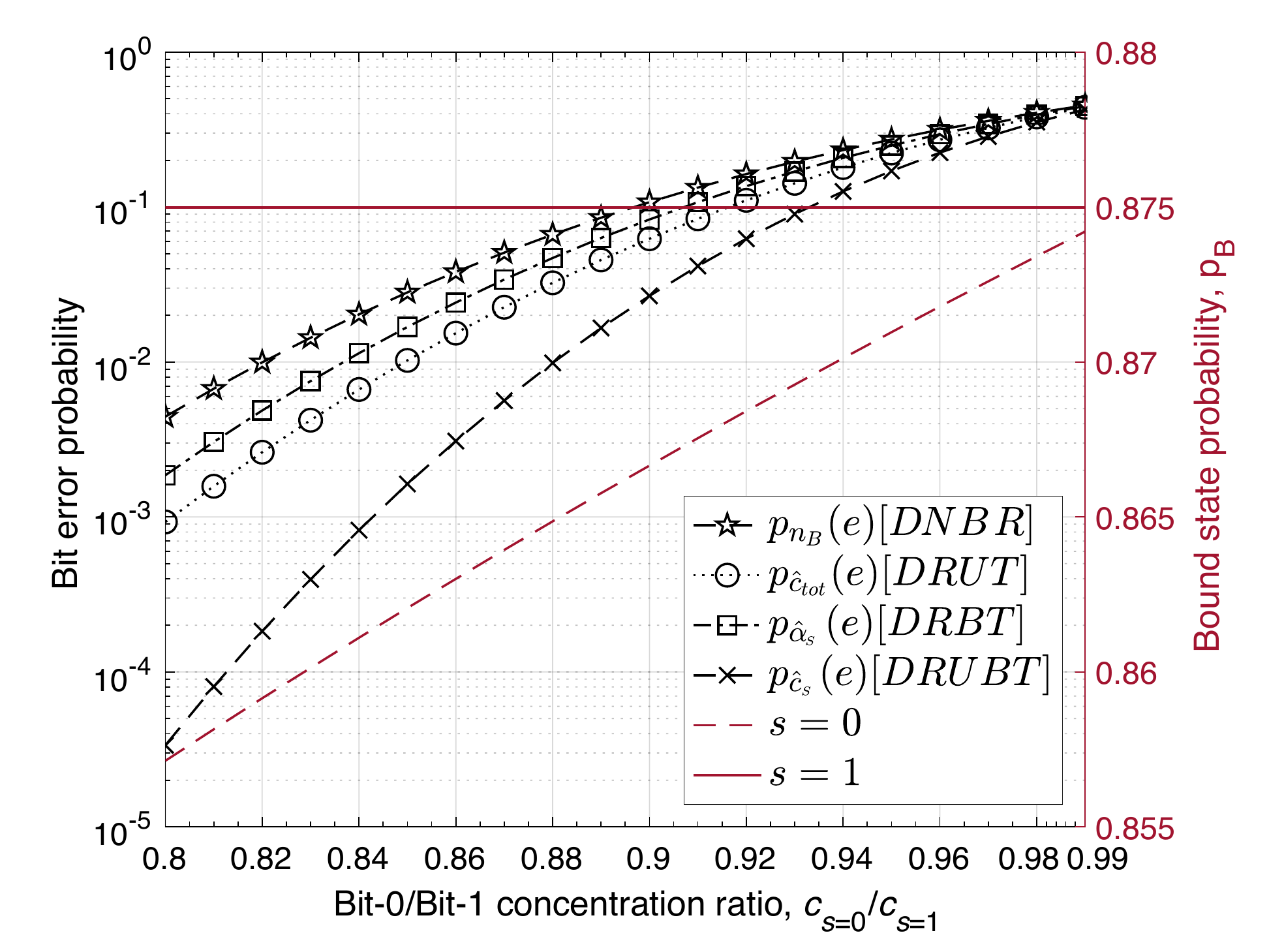}
		\label{fig:Bit0ConcentrationClose}
	}
	\caption{(a) Bit error probability with varying ratio of the received concentrations corresponding to bit-0 and bit-1 transmissions, i.e., $c_{s=0}/c_{s=1}$. A magnified view is provided in (b).}
	\label{fig:Bit0ConcentrationMain}
\end{figure*}
\subsection{Effect of Bit-0/Bit-1 Concentration Ratio}
We also analyze the effect of the distance between bit-0 and bit-1 in terms of the received concentration of information molecules. In the case of high ISI in the MC channel, it is very likely that a considerable amount of information molecules from previous transmissions remains in the reception space. As a result, the ratio between the distinct concentration levels corresponding to bit-0 and bit-1 transmissions may approach 1, obstructing the discrimination between them. Here we gradually change the ratio between the received information molecule concentrations for bit-0 and  bit-1 from 0.1 to 0.99, and provide the results in Fig. \ref{fig:Bit0Concentration} (with a magnified version in Fig. \ref{fig:Bit0ConcentrationClose}). As the concentration ratio approaches 1, all the detection methods fail to provide an acceptable error performance. On the other hand, DRUBT performs significantly better than any of the other detection methods tested. It is also to be noted that the performance of DRUT becomes the worst among all, when the concentration levels are well-separated, e.g., in the case of very low-level ISI. In this particular range, DNBR, which is the simplest detection method, performs very well. This indicates that even in the presence of interferer molecules, the instantaneous number of bound receptors can provide sufficient statistics for detection as long as the concentration levels for bit-0 and bit-1 are well-separated, and the interferer concentration is at a moderate level.

\begin{figure}[!t]
	\centering
	\includegraphics[width=8cm]{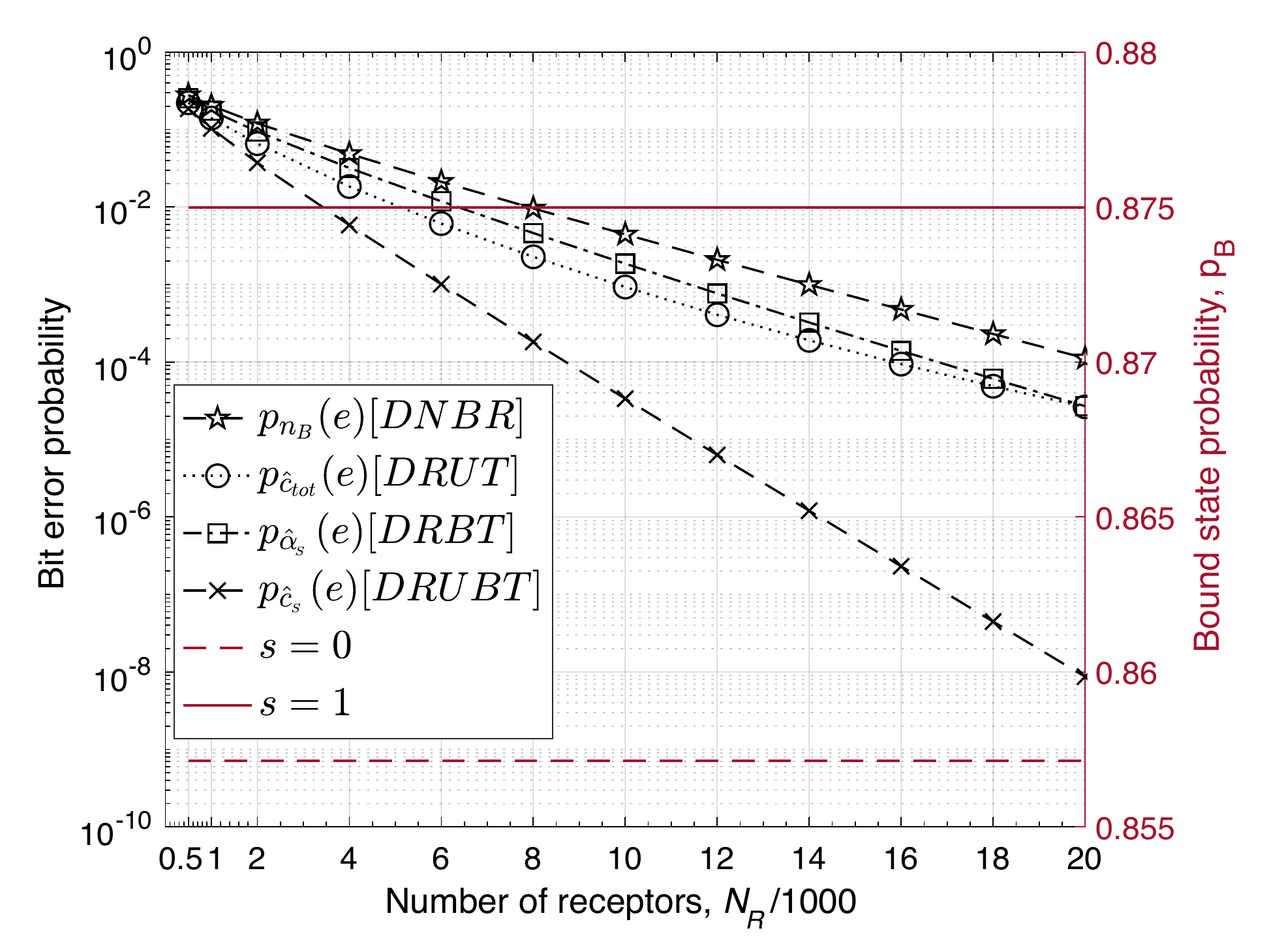}
	\caption{Bit error probability as a function of number of receptors. }
	\label{fig:NumberofSamples}
\end{figure}

\subsection{Effect of Number of Receptors}
The number of receptors determines the number of independent samples taken for detection. For example, in DNBR, the instantaneous number of bound receptors is the sum of $N_R$ random variables independently following Bernoulli distribution. In DRUT and DRUBT, the total unbound time $T_u$ is the sum of $N_R$ unbound time intervals which independently follow exponential distributions. In DRBT and DRUBT, each of the independent receptors is assumed to sample only one binding event, leading to the observation of $N_R$ independent binding events. All of the investigated detection methods show similar, almost log-linear, performance trends with the varying number of receptors, as shown in Fig. \ref{fig:NumberofSamples}. Nevertheless, the most significant performance improvement is observed with DRUBT as it relies on both unbound time and bound time statistics taken independently from all receptors.


\section{Discussion on Implementation} 
\label{sec:implementation}
The introduced detection methods are practical in the sense that they can be implemented by biologically plausible synthetic receptors and CRNs in MC receivers. In this section, we discuss four different receptor designs that can transduce the required receptor statistics (i.e., number of bound receptors, total unbound time, number of binding events with the durations within specific time intervals) into the concentration of intracellular molecules, i.e., secondary messengers. The receptor designs incorporate an activation mechanism, which was previously introduced in \cite{kuscu2019channel}, to control the start time and the duration of the sampling process. We also discuss potential CRNs, which can chemically process the generated secondary messengers in order to perform the analog and digital computations required for the detection. Lastly, we provide a discussion of the state-of-the-art synthetic biology tools and emerging research trends that can enable the implementation of the proposed receptors and CRNs.

\subsection{Activation Mechanism}
In order to control the start time and the duration of the sampling, the receiver must have an activation mechanism that allows the receptors to generate secondary messengers only when they are in the active state. In \cite{kuscu2019channel}, we proposed an activation mechanism that enables the sampling of the required statistics from independent receptors only once during each sampling period. The synthetic receptor designs,  which will be introduced next, incorporate this mechanism. In this scheme, the cell generates activator molecules $A^+$, which can rapidly diffuse and interact only with the inactive receptors at the reaction rate $\omega$, and shift them to active or intermediate states depending on the adopted receptor design. The generation of the activation molecules is governed by the following reaction
\begin{equation}
\ce{ {\varnothing}  ->[{g(t)\psi^+}] {A^+}},
\end{equation}
where $g(t) \psi^+$ is the time-varying generation rate of $A^+$ molecules, with $g(t) \approx \delta(t-t_A)$ being a very short pulse signal centered around the activation time $t_A$. The generation of $A^+$ molecules is followed by 
the generation of deactivation molecules $A^-$, i.e.,
\begin{equation}
\ce{ {\varnothing}  ->[{d(t)\psi^-}] {A^-}}.
\end{equation}
The generation rate of deactivation molecules is given by $d(t) \psi^-$. Here $d(t) \approx \delta(t-t_D)$ is an impulse-like signal centered around the deactivation time $t_D$. The activation molecules are degraded by the deactivation molecules with the reaction rate $\rho$, i.e.,
\begin{equation}
\ce{ {A^+ + A^-} ->[{\rho}] {\varnothing} }.
\end{equation}
In this way, the duration of the overall sampling process is controlled by the receiver cell. Note that the reaction rates governing the activation mechanism, i.e., $\psi^+$, $\psi^-$, $\omega$, $\rho$, should be very high compared to the ligand-receptor binding/unbinding reaction rates to prevent the inactivated receptors from being re-activated in the same sampling interval. 
\begin{figure*}[!t]
	\centering
	\includegraphics[width=16cm]{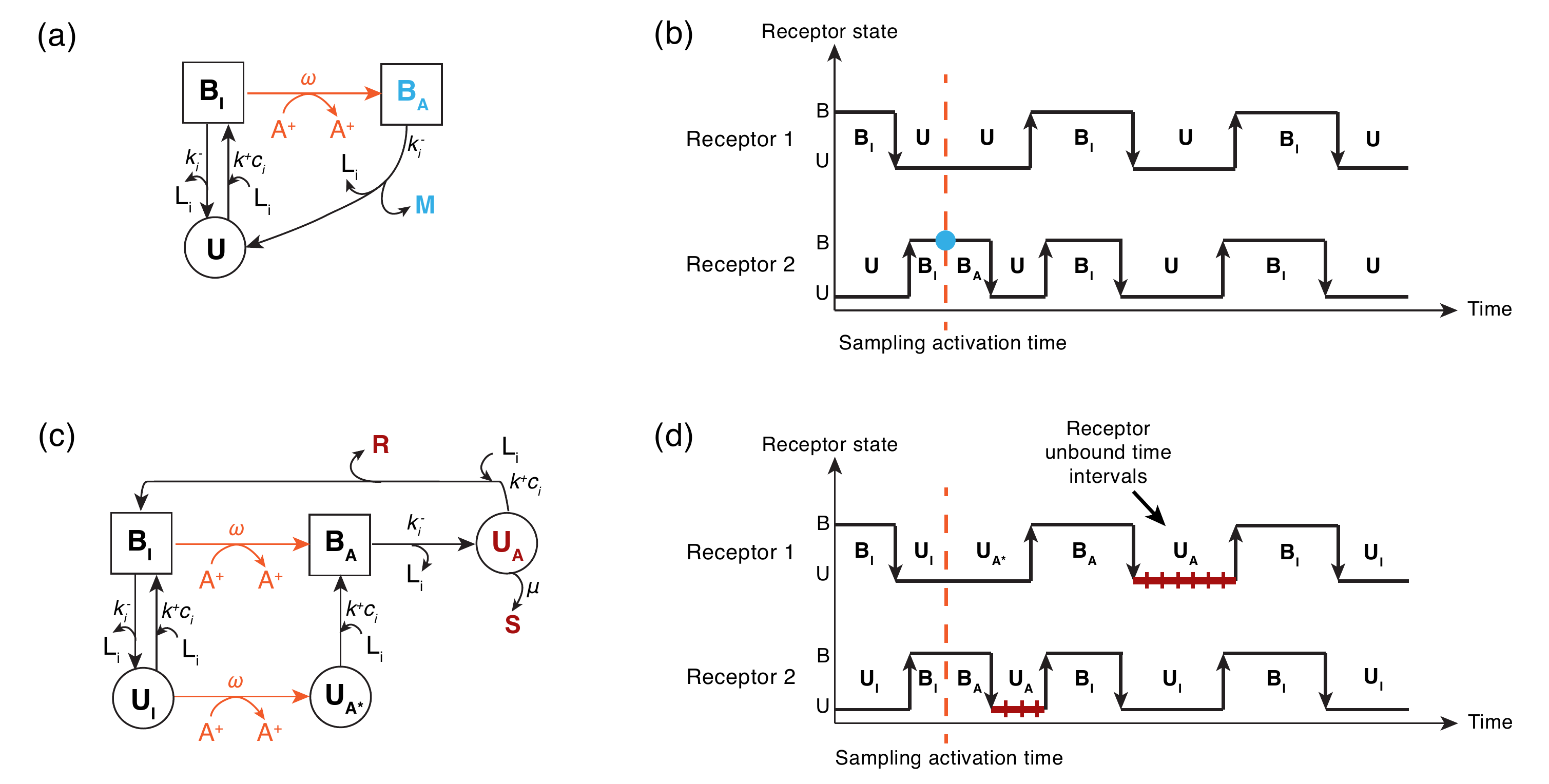}
	\caption{(a) Receptor design for DNBR. (b) Sampling of the number of bound receptors. (c) Receptor design for DRUT. (d) Sampling of the receptor unbound time intervals. }
	\label{fig:implementation1}
\end{figure*}


\subsection{Implementation of DNBR}
In DNBR, we need a representation of number of bound receptors at the sampling time in terms of concentration of intracellular molecules. A potential synthetic receptor design is provided in Fig. \ref{fig:implementation1}(a). In this design, the receptor has three states, unbound state $\U$, inactive bound state $\B_I$, and active bound state $\B_A$. The receiver cell releases intracellular activation molecules at the time of sampling. The released activation molecules $A^+$ only react with $\B_I$, and convert them to $\B_A$. The active bound receptors release an intracellular molecule $M$ upon unbinding from a ligand, and returns to the unbound state. As a result, the number of $M$ molecules encodes the number of bound receptors at the sampling time. This intracellular signal can also be amplified if the receptors are designed to release multiple $M$ molecules in the active state. 

The next process is to compare the intracellular concentration of $M$ molecules to a threshold encoded by a different secondary messenger. A simple comparator can be implemented through the following reaction
\begin{equation}
\ce{ {M + X} ->[{\xi}] {\varnothing} }, 
\end{equation}
where $X$ molecules encode the threshold given in \eqref{threshold3} for DNBR. If any $M$ molecule remains inside the cell after this reaction, the receiver decides bit-1, otherwise it decides bit-0. In the case that the intracellular concentration of $M$ molecules is amplified, the threshold signal should be also amplified proportionally.  



\subsection{Implementation of DRUT}
DRUT requires the transduction of the total unbound time of receptors into the concentration of second messengers. We propose a synthetic receptor design with an activation mechanism demonstrated in Fig. \ref{fig:implementation1}(c), to guarantee that only one unbound time information is acquired from each independent receptor. In this design, the receptor has 5 states: inactive unbound ($\U_I$), intermediate unbound ($\U_A^\ast$), active unbound ($\U_A$), inactive bound ($\B_I$), and active bound ($\B_A$) states. At the sampling time, the receiver releases the activation molecules, which can rapidly diffuse and react only with the receptors at $\U_I$ or $\B_I$ states. If the receptor is already unbound at the time of activation, it stays idle until the next complete unbound interval, when it starts releasing intracellular molecules that encode the unbound time duration. Therefore, an already-unbound receptor first goes into the intermediate state $\U_A^\ast$ upon receiving the activation signal, which is followed by $\B_A$ and $\U_A$ states. If it is bound when activated, the next unbinding event takes it to the active unbound state $\U_A$. Receptors at $\U_A$ state release intracellular molecules $S$ at a fixed rate, and upon the first binding event, they transition to the inactive bound state $\B_I$, simultaneously releasing a single molecule of a different type $R$ in order to encode the number of independent samples taken.

The resulting concentration of intracellular $S$ molecules encodes the total unbound time of receptors over a single period of sampling. These second messengers together with $R$ molecules can be processed by a CRN, which biochemically implements the total concentration estimator, given in Eq. \eqref{ctot_estimator}. An example CRN could be as follows: 
\begin{equation}
\ce{ {R} ->[{1}] {R} + { Y} },
\end{equation}
\begin{equation}
\ce{ S + Y ->[{k^+}] S }.
\end{equation}
In this CRN, we introduce another type of intracellular molecule $Y$, which is produced by $R$ molecules at the unit rate, while consumed by $S$ molecules at the common binding rate of ligands $k^+$. The rate equation of this CRN can be written as
\begin{align}
\frac{d\E[ n_{Y}] }{dt} = \E[n_{R}]  - k^+ \E[n_{S}]  \E[ n_{Y}], 
\end{align}
where $n_Y$, $n_R$, and $n_S$ are the number of $Y$, $R$, and $S$ molecules, respectively. Given the initial condition $\E[ n_{Y}^0] = 0$, the steady-state solution for $\E[ n_{Y}]$ can be obtained as
\begin{align}
\E[ n_{Y}^{ss}] = \frac{\E[n_{R}^{ss}]}{k^+ \E[n_{S}^{ss}]}, 
\end{align}
Given that $n_R^{ss}$ and $n_S^{ss}$ encode the number of independent receptors and the total unbound time, respectively, the resulting number of $Y$ molecules at steady-state $n_Y^{ss}$ approximates the total concentration estimator $\hat{c}_{tot} = \frac{N-1}{k^+ T_u}$.

For decision, the comparator reaction utilized in DNBR can also be implemented here, i.e., 
\begin{equation}
\ce{ {Y + X} ->[{\xi}] {\varnothing} },
\end{equation}
where the number of $X$ molecules encodes the optimal threshold value. If any $Y$ molecule remains in the cell as a result of this reaction, the receiver decides bit-1, otherwise it decides bit-0.

%
%
%
%

\subsection{Implementation of DRBT}
\begin{figure*}[!t]
	\centering
	\includegraphics[width=16cm]{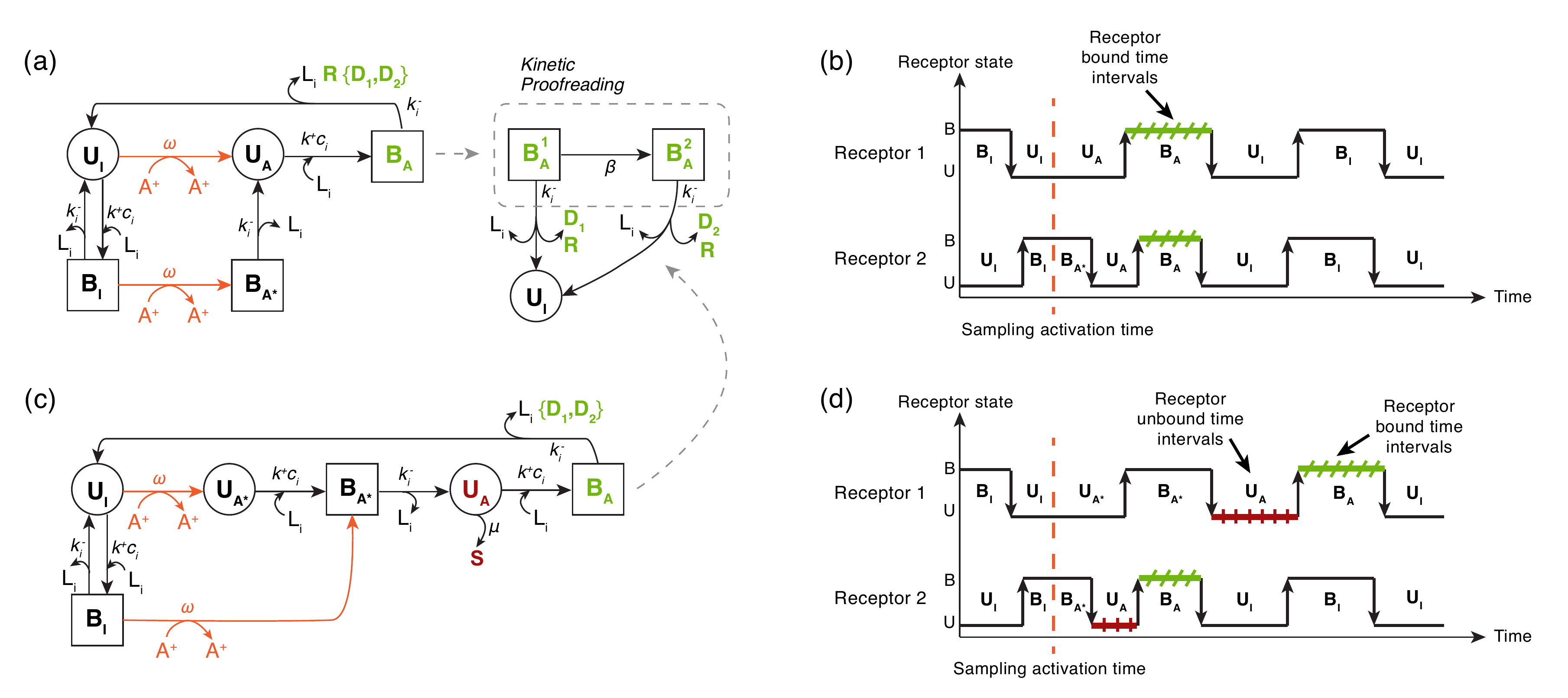}
	\caption{(a) Receptor design for DRBT. (b) Sampling of the number of binding events of durations within specific time intervals. (c) Receptor design for DRUBT. (d) Sampling of the receptor unbound time intervals and the number of binding events of durations within specific time intervals. }
	\label{fig:implementation2}
\end{figure*}

DRBT requires the number of binding events with durations that fall into specific time ranges to be encoded into the concentration of intracellular molecules. Following our proposal in \cite{kuscu2019channel}, we introduce a synthetic receptor design with an activation mechanism and a modified KPR mechanism, as demonstrated in Fig. \ref{fig:implementation2}(a). The activation mechanism is similar to those introduced for DNBR and DRUT, and ensures that only one binding time information is received from each independent receptor, as shown in Fig. \ref{fig:implementation2}(b). In this design, when an active unbound receptor $\U_A$ binds to a ligand, it switches to the active bound state $\B_A$ where the KPR mechanism is activated.

The KPR mechanism consists of two substates, $\B_A^1$ and $\B_A^2$, with a unidirectional state transition rate $\beta$, which can be set as a function of the time threshold $T_1$ as follows  
\begin{align}
\beta = \kappa_i/T_1 ,
\end{align}
where $\kappa_i$ is a tuning parameter adopted to optimize the transition rate for accurate sampling of the receptor bound time durations. Our analysis in \cite{kuscu2019channel} shows that $\kappa_i = 3/5$ provides reasonable accuracy in representing the number of binding events $n_{b,1}$ and $n_{b,2}$ with second messengers via the stochastic KPR mechanism.

A receptor is allowed to return to the inactive unbound state $\U_I$ at any time by unbinding from the bound ligand. While returning to the state $\U_I$, the receptor releases a single $R$ molecule encoding the number of independent samples, and one of the intracellular molecules $D_1$ or $D_2$, depending on the last visited KPR substate. In this way, the KPR mechanism allows to discriminate long binding events, which are more likely to be resulting from the molecules with higher affinity, from short binding events through encoding the number of corresponding binding events into the number of $D_1$ and $D_2$ molecules. A steady-state analysis of a similar KPR mechanism is provided in \cite{kuscu2019channel}. 

The generated intracellular molecules $R$, $D_1$ and $D_2$ are biochemically processed by a CRN to realize the ratio estimator $\hat{\alpha}_s$, given in \eqref{alpha_estimator}. An example CRN can be as follows
\begin{equation}
\ce{ {D_1} ->[{w_{2,1}}] {D_1} + { Y} },
\end{equation}
\begin{equation}
\ce{ {D_2} ->[{w_{2,2}}] {D_2} + { Y} },
\end{equation}
\begin{equation}
\ce{ R + Y ->[{1}] R }.
\end{equation}
In this CRN, the intracellular molecules $Y$ are produced by $D_1$ and $D_2$ with the reactions rates $w_{2,1}$ and $w_{2,2}$, respectively. They are consumed by $R$ molecules at the unit rate. The rate equation of this CRN can then be written as 
\begin{align}
\frac{d\E[ n_{Y}] }{dt} = w_{2,1} \E[n_{D_1}] + w_{2,2} \E[n_{D_2}] -  \E[n_{R}]  \E[ n_{Y}].
\end{align}
Given the initial condition $\E[ n_{Y}^0] = 0$, the steady-state solution for $\E[ n_{Y}]$ can be obtained as follows
\begin{align}
\E[ n_{Y}^{ss}] = \frac{w_{2,1} \E[n_{D_1}^{ss}] + w_{2,2} \E[n_{D_2}^{ss}] }{E[n_{R}^{ss}]}. 
\end{align}
If $n_{D_1}^{ss}$ and $n_{D_2}^{ss}$ encode $n_{b,1}$ and $n_{b,2}$, respectively, and $n_{R}^{ss}$ encodes the number of independent samples $N$, then the number of $Y$ molecules at steady-state $n_{Y}^{ss}$ approximates the concentration ratio estimator $\hat{\alpha}_s = \left( n_{b,1} w_{2,1} + n_{b,2} w_{2,2} \right)/N.$

The comparator reaction for decoding the transmitted bit can be realized in a similar way as DNBR and DRUT, i.e., 
\begin{equation}
\ce{ {Y + X} ->[{\xi}] {\varnothing} },
\end{equation}
where the number of $X$ molecules encodes the optimal threshold given in \eqref{threshold3}. If any $Y$ molecule remain in the cell as a result of this reaction, the receiver decides bit-1, otherwise bit-0 is decided. 

%
%
%

\subsection{Implementation of DRUBT}

DRUBT requires the transduction of both the total unbound time, and the number of binding events of durations within specific time ranges. Hence, the combination of the synthetic receptor designs introduced for DRUT and DRBT enables the required functionality. The receptor design given in Fig. \ref{fig:implementation2}(c) ensures that only a single pair of complete unbound and bound time duration is sampled from each independent receptor. For DRUBT, receptors are not required to generate $R$ molecules, because the concentration estimator is not a function of the number of independent samples.

Given that the number of generated $S$, $D_1$ and $D_2$ molecules encodes $T_U$, $n_{b,1}$ and $n_{b,2}$, respectively, the CRN for the concentration estimator $\hat{c}_s$, given in \eqref{eq:csestimator}, can be implemented as follows 
\begin{equation}
\ce{ {D_1} ->[{w_{2,1}}] {D_1} + { Y} },
\end{equation}
\begin{equation}
\ce{ {D_2} ->[{w_{2,2}}] {D_2} + { Y} },
\end{equation}
\begin{equation}
\ce{ {S} + {Y} ->[{k^+}] S }.
\end{equation}
The rate equation of this CRN can be written as
\begin{align}
\frac{d\E[ n_{Y}] }{dt} = w_{2,1} \E[n_{D_1}] + w_{2,2} \E[n_{D_2}] -  k^+ \E[n_{S}]  \E[ n_{Y}].
\end{align}
Given the initial condition $\E[ n_{Y}^0] = 0$, the steady-state solution for $\E[ n_{Y}]$ can be obtained as
\begin{align}
\E[ n_{Y}^{ss}] = \frac{w_{2,1} \E[n_{D_1}^{ss}] + w_{2,2} \E[n_{D_2}^{ss}] }{k^+ E[n_{S}^{ss}]}. 
\end{align}
As is obvious, the number of $Y$ molecules at steady-state $n_{Y}^{ss}$ approximates $\hat{c}_s =  \frac{1}{k^+ T_u}  \left( n_{b,1} w_{2,1} + n_{b,2} w_{2,2} \right)$ for  $N \gg 1$.

The following comparator can be applied here as well for decoding:  
\begin{equation}
\ce{ {Y + X} ->[{\xi}] {\varnothing} },
\end{equation}
where the number of $X$ molecules encodes the optimal threshold value given in \eqref{threshold3}. If eventually $Y$ molecules outnumber $X$ molecules, the receiver decides bit-1, otherwise it decides bit-0.

\subsection{Discussion}
	The receptor and CRN designs proposed for the implementation of the MC detection methods are biologically plausible in the sense that similar designs are already utilized by living cells, and synthetic receptors and intracellular CRNs obtained via modification of these natural designs or via de novo designs are becoming increasingly sophisticated with the recent advances in synthetic biology. 


The introduced detection methods rely on multi-state receptors with the state transitions being realized by binding/unbinding of ligands or activation/deactivation molecules. Each of these states either determines the function of the receptors, e.g., releasing a type of intracellular molecules, or acts as an intermediate state regulating the set of next feasible state transitions. Such multi-state receptors are widely utilized in biological cells \cite{wong2020tcrbuilder, lau2013conformational}. In these natural systems, the receptor states correspond to different conformational states of the receptors, where the receptors typically manifest different binding/signaling characteristics \cite{kahsai2011multiple}. 

Our receptor designs in DRBT and DRUBT also incorporate a modified KPR mechanism to discriminate between the binding events of the information and the interferer molecules. KPR has long been speculated to be the mechanism underlying the impressive performance of T cells in discriminating between the lower-affinity self-ligands and higher-affinity antigens to evoke the immune response \cite{mckeithan1995kinetic, rabinowitz1996kinetic}, and recently a strong empirical evidence has been provided to corroborate this hypothesis \cite{yousefi2019optogenetic}. In such KPR systems, intermediate receptor states delay the activation of the receptor and consequent release of intracellular molecules as a way of exploiting the statistical difference between binding durations of self-ligands and antigens. The modified KPR scheme in our receptor designs only slightly differs from the natural KPR scheme in the sense that the unbinding of ligands during the receptors' first KPR substate (intermediate state) also results in the release of an intracellular molecule (i.e., D$_1$ molecules in Fig. \ref{fig:implementation2}).


Various biosensing and therapeutic applications have already been developed with engineered cells through such synthetic modifications over natural receptor systems \cite{chang2020synthetic, hicks2020synthetic}. This progress has been particularly fueled by the introduction of novel design frameworks (e.g., Tango assay \cite{barnea2008genetic}, Modular Extracellular Sensor Architecture (MESA) \cite{daringer2014modular}, SynNotch \cite{morsut2016engineering}) for modular receptors that combine diverse components of natural receptors, e.g., G protein-coupled receptors (GPCRs), Notch receptors, TCRs, to enable new receptor functions. For example, binding affinity and kinetic rates of the receptors can now be tuned to favor the binding of specific types of extracellular and intracellular ligands \cite{chervin2008engineering, bowerman2009engineering}. These synthetic receptors can also be seamlessly coupled to the orthogonal intracellular signaling pathways to multiplex the biosensing \cite{chen2021programmable}. Moreover, multiple ligand inputs can be AND-gated through synthetic receptors to realize combinatorial sensing \cite{roybal2016precision}. As a practical example, the specificity of genetically modified T cells has been redirected to tumor-associated antigens with chimeric antigen receptors (CARs) to improve their therapeutic efficacy \cite{jena2010redirecting}. More importantly for the receptor designs proposed in this paper, the progress in computational de novo protein design and engineering allows the creation of arbitrary protein conformational states with tailored interaction parameters, which can be translated to multi-state synthetic receptors \cite{huang2016coming, pan2021recent, quijano2021novo, wei2020computational, dagliyan2013rational}.


In parallel to the advances in de novo protein engineering, the research in synthetic biology of signaling networks, which are abstracted as CRNs, has been witnessing an exciting shift towards synthetic post-translational protein circuits from more conventional synthetic genetic circuits \cite{chen2021programmable}. Post-translational protein circuits are much faster than their genetic counterparts, which involve the slow transcription and translation processes \cite{chen2021programmable}. They can be more directly and rapidly linked to the protein-based receptor systems in a seamless manner, enabling receptor-integrated CRN-based computations in synthetic cells \cite{gao2018programmable}. Moreover, protein circuits reinforced with de novo protein designs offer a higher degree of orthogonality in intracellular reaction pathways, enabling the implementation of more diverse and parallel functions within synthetic cells \cite{chen2021programmable, chen2019programmable, lim2010designing}. Therefore, protein circuits are promising for the design of more sophisticated CRNs, which rely on protein interactions, e.g., binding, cleaving, and chemical modification, with the interaction rates tuned specifically to implement the desired arithmetic and logic operations. 

In summary, we believe that the ongoing progress in synthetic biology reinforced with the recently introduced receptor design frameworks and the emerging branch of protein engineering will allow the implementation of the proposed biologically-plausible multi-state synthetic receptors, receptor-integrated CRNs, and the overall MC detection systems in living cells in near future.

\section{Conclusion}
\label{sec:conclusion}
In this paper, we investigate the performance of four different MC detection methods in the case of interference resulting from molecules having similar binding affinity as the information-carrying molecules. The detection methods are based on different statistics of the ligand-receptor binding reaction, e.g., instantaneous number of bound receptors, duration of receptors' bound and unbound time intervals, that reveal information about the concentration and binding affinity of the molecules. The methods vary in complexity; however, they are all biologically plausible, and we believe that the ongoing progress and sophistication in synthetic biology, and particularly in de novo protein engineering,  will soon enable the practical implementation of the required synthetic receptors and CRNs in biological MC devices. Our analyses show that the effect of molecular interference on the detection performance can be substantially reduced by using the combination of unbound and bound time durations of receptors instead of relying solely on the number of bound receptors, which has been the prevalent approach in the previous literature.

	\begin{figure}[!t]
	\centering
	\includegraphics[width=7cm]{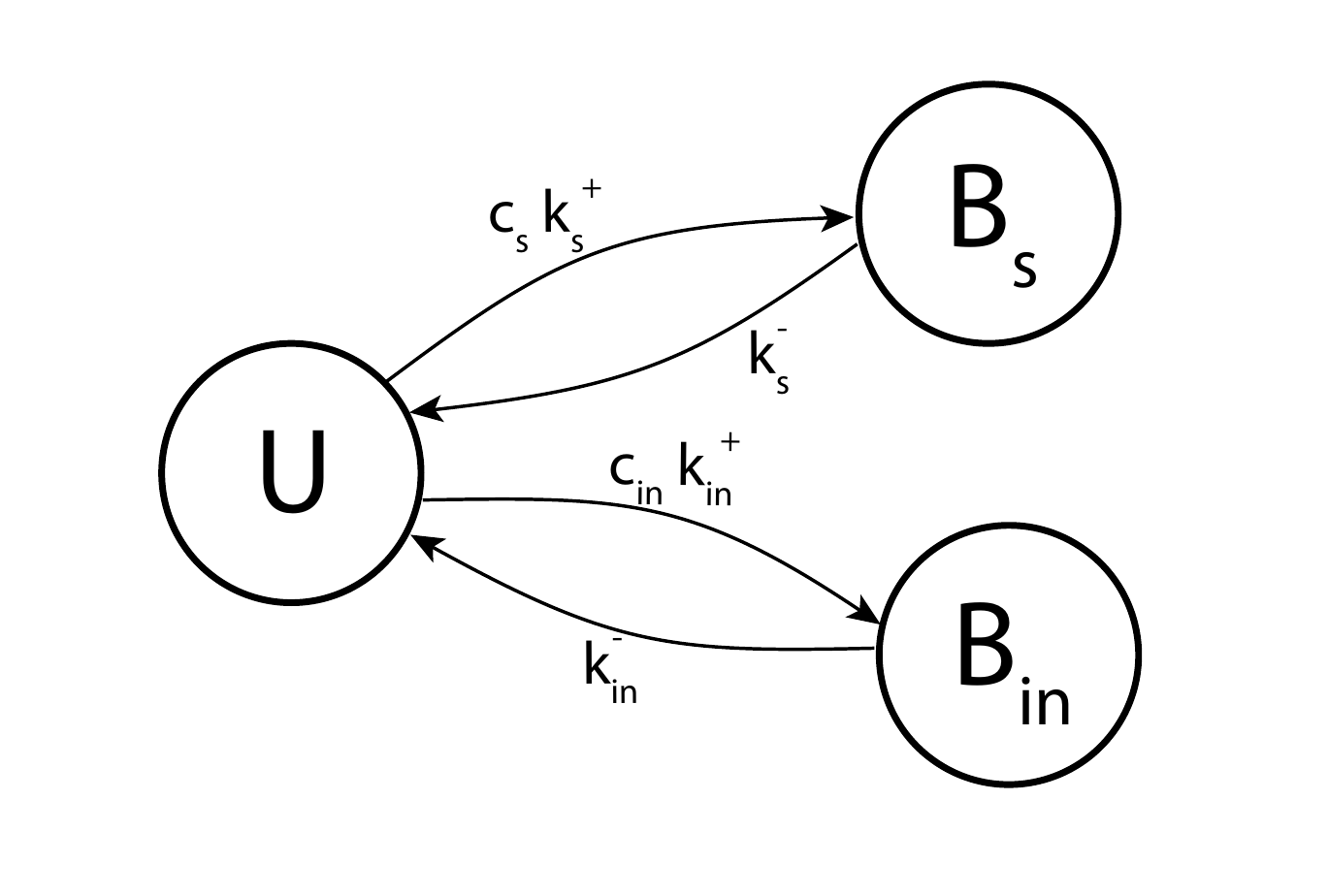}
	\caption{State diagram of the CTMP model of receptor binding process.}
	\label{fig:Markov}
\end{figure}

\appendices
	\section{Bound State Probability of Receptors at Equilibrium in the Presence of Interferer Molecules}
	\label{AppendixA}
	In the presence of interferer molecules with concentration $c_{in}$, the receptor binding process can be represented by a 3-state CTMP, with the unbound state (U) accompanied by two bound states, B$_s$ and B$_{in}$, corresponding to the binding of information and interferer molecules, respectively. The state diagram of this CTMP along with the corresponding state transition rates is shown in Fig. \ref{fig:Markov} .
	The rate matrix is then constructed as follows
	\renewcommand{\kbldelim}{(}
	\renewcommand{\kbrdelim}{)}
	\[
	\bm{R} = \kbordermatrix{
		& \U & \B_{s} & \B_{in}  \\
		\U & -(k_s^+ c_s + k_{in}^+ c_{in}) & k_s^+ c_s  & k_{in}^+ c_{in} \\
		\B_{s} & k_s^- & -k_s^- & 0  \\
		\B_{in} & k_{in}^- & 0 & -k_{in}^- 
	}
	\]

The equilibrium probabilities of the receptor states $\bm{\theta} = [\p_\U ~~ \p_{\B_s} ~~ \p_{\B_{in}}]$ can be obtained by solving the following linear equations, $\bm{\theta} \bm{R} = 0$, and $\bm{\theta} \bm{e} = 1$, where $\bm{e}$ is all-ones vector \cite{whitt2006continuous}. These equations represent, respectively, the detailed balance condition at equilibrium, and the fact that the sum of all probabilities must be equal to 1. From this system of linear equations, we can obtain the unbound state probability at equilibrium as 
	\[
	\p_\U  = \frac{k_s^- k_{in}^-}{k_s^+ k_{in}^- c_s + k_s^- k_{in}^+ c_{in} + k_s^- k_{in}^-}.
	\]
	The bound state probability of the receptors at equilibrium $\p_\B = \p_{\B_s} + \p_{\B_{in}}$ is then given by
	\[
	\p_\B  = 1 - \p_\U = \frac{k_s^+ k_{in}^- c_s + k_s^- k_{in}^+ c_{in}}{k_s^+ k_{in}^- c_{s} + k_s^- k_{in}^+ c_{in} + k_s^- k_{in}^-}. 
	\]
	Substituting $K_D = k^-/k^+$, we recover Eq. \eqref{probBinding2}:
	\[
	\p_\B = \frac{c_s/K_D^s + c_{in}/K_D^{in}  }{1 + c_s/K_D^s + c_{in}/K_D^{in}}.
	\]

\bibliographystyle{ieeetran}
\bibliography{DetectionUnderInterferenceArxivRevised}
\end{document}